\documentclass[AMS,STIX1COL]{WileyNJD-v2}

\articletype{Article Type}%

\received{<day> <Month>, <year>}
\revised{<day> <Month>, <year>}
\accepted{<day> <Month>, <year>}

\usepackage{hyphenat}
\usepackage{lineno}
\usepackage{amsmath,amssymb,amsfonts,amsthm}
\usepackage{bbm}
\usepackage{physics}
\usepackage{xcolor}
\usepackage{psfrag}
\usepackage{algorithm}
\usepackage{subfigure}

\newcommand{\mb}[1]{\ensuremath{\mathbf{#1}}}
\newcommand{\ctensor}{\mathbbm{C}}
\newcommand{\as}{\alpha_{\text{s}}}
\newcommand{\am}{\alpha_{\text{m}}}
\newcommand{\Xalpha}{\mathrm{X}_{\alpha}}
\newcommand{\Xb}{\mathrm{X}_{\beta}}
\newcommand{\Xm}{\mathrm{X}_{\am}}
\newcommand{\Xa}{\mathrm{X}_{\alpha_{\text{s}}}}

\newcommand{\Xaeq}{\mathrm{X}_{\alpha}^{\text{eq}}}
\newcommand{\dT}{\dot{T}}
\newcommand{\dTm}{\dot{T}_{\alpha_{\text{m}}, \text{min}}}
\newcommand{\Tbs}{T_{\alpha_s,\text{end}}}
\newcommand{\Tbe}{T_{\alpha_s,\text{sta}}}
\newcommand{\krateeq}{\mathrm{k}_{\ao}^{\text{eq}}}
\newcommand{\ao}{\alpha}
\newcommand{\Xmeqo}{\mathrm{X}_{\am,0}^{\text{eq}}}
\newcommand{\Xmeq}{\mathrm{X}_{\am}^{\text{eq}}}
\newcommand{\kratem}{\mathrm{k}_{\am}^{\text{eq}}}
\newcommand{\Tms}{T_{\alpha_{\text{m}},\text{sta}}}
\newcommand{\dXa}{\dot{\mathrm{X}}_{\alpha_{\text{s}}}}
\newcommand{\dXm}{\dot{\mathrm{X}}_{\am}}
\newcommand{\dXb}{\dot{\mathrm{X}}_{\beta}}
\newcommand{\dXbtoa}{\dot{\mathrm{X}}_{\beta\rightarrow\as}}
\newcommand{\dXmtoa}{\dot{\mathrm{X}}_{\am\rightarrow\as}}
\newcommand{\dXmtob}{\dot{\mathrm{X}}_{\am\rightarrow\beta}}
\newcommand{\dXatob}{\dot{\mathrm{X}}_{\as\rightarrow\beta}}
\newcommand{\dXbtoam}{\dot{\mathrm{X}}_{\beta\rightarrow\am}}

\begin{document}

\title{{Physics-Based Modeling and Predictive Simulation of Powder Bed Fusion Additive Manufacturing Across Length Scales}}

\author[1]{Christoph Meier*}

\author[1,2]{Sebastian L. Fuchs}

\author[1]{Nils Much}

\author[1]{Jonas Nitzler}

\author[3]{Ryan W. Penny}

\author[1]{Patrick M. Praegla}

\author[1]{Sebastian D. Pr\"oll}

\author[1]{Yushen Sun}

\author[1,3]{Reimar Weissbach}

\author[1,4]{Magdalena Schreter}

\author[5]{Neil E. Hodge}

\author[3]{A. John Hart}

\author[1]{Wolfgang A. Wall}

\authormark{Meier \textsc{et al.}}

\address[1]{\orgdiv{Institute for Computational Mechanics}, \orgname{Technical University of Munich}, \orgaddress{\state{Munich}, \country{Germany}}}

\address[2]{\orgdiv{Institute for Continuum and Material Mechanics}, \orgname{Hamburg University of Technology}, \orgaddress{\state{Hamburg}, \country{Germany}}}

\address[3]{\orgdiv{Department of Mechanical Engineering}, \orgname{Massachusetts Institute of Technology}, \orgaddress{\state{Cambridge}, \country{USA}}}

\address[4]{\orgdiv{Unit of Strength of Materials and Structural Analysis}, \orgname{University of Innsbruck}, \orgaddress{\state{Innsbruck}, \country{Austria}}}

\address[5]{\orgdiv{Methods Development Group}, \orgname{Lawrence Livermore National Laboratory}, \orgaddress{\state{Livermore}, \country{USA}}}

\corres{*Christoph Meier, Institute for Computational Mechanics, Technical University of Munich, Boltzmannstrasse 15, 85748, Garching, Germany. \email{meier@lnm.mw.tum.de}}


\abstract[Summary]{Powder bed fusion additive manufacturing (PBFAM) of metals has the potential to enable new paradigms of product design, manufacturing
and supply chains while accelerating the realization of new technologies in the medical, aerospace, and other industries. Currently, wider adoption of PBFAM is held back by difficulty in part qualification, high production costs and low production rates, as extensive process tuning, post-processing, and inspection are required before a final part can be produced and deployed. Physics-based modeling and predictive simulation of PBFAM offers the potential to advance fundamental understanding of physical mechanisms that initiate process instabilities and cause defects. In turn, these insights can help link process and feedstock parameters with resulting part and material properties, thereby predicting optimal processing conditions and inspiring the development of improved processing hardware, strategies and materials. This work presents recent developments of our research team in the modeling of metal PBFAM processes spanning length scales, namely mesoscale powder modeling, mesoscale melt pool modeling, macroscale thermo-solid-mechanical modeling and microstructure modeling. Ongoing work in experimental validation of these models is also summarized. In conclusion, we discuss the interplay of these individual submodels within an integrated overall modeling approach, along with future research directions.}



\maketitle


\section{Introduction} \label{sec:intro}

\noindent
Currently, widespread adoption of metal powder bed fusion additive manufacturing (PBFAM) processes such 
as selective laser melting (SLM) or electron beam melting (EBM) is mainly held back by reliable part qualification,
high production costs and low production rate. Usually extensive manual process tuning via trial and error, post-processing, and inspection are required to ensure compliance with specified quality requirements and defect thresholds before a final part can be produced and deployed. The physics-based modeling and predictive simulation of these processes offers the potential to advance fundamental understanding of the governing process physics on different length scales and mechanisms of process instability and 
defect creation, which link process and feedstock parameters with resulting part and material properties.

\begin{figure}[h!!!]
\begin{centering}
\includegraphics[width=0.95\textwidth]{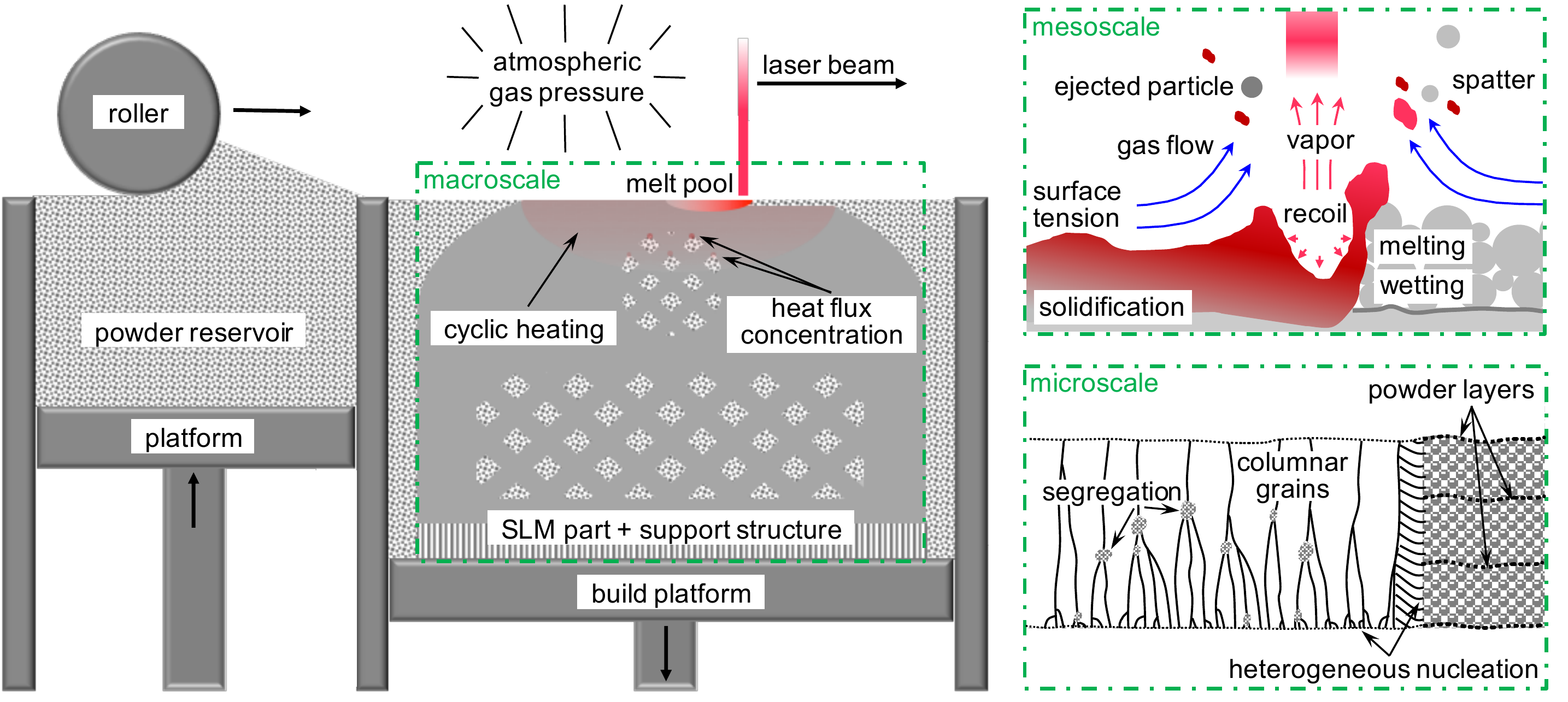}
\caption{{Setup of SLM process and schematic visualization of macroscale, mesoscale and microscale view~\cite{Meier2017}.}}
\label{fig:Fig_zy1}
\end{centering}
\end{figure}

The multiscale nature of PBFAM processes requires the use of individual models tailored to study physical phenomena occurring on these different length scales. Thus, existing models are typically classified in macroscale, mesoscale and microscale approaches~\citep{Meier2017} (see Figure~\ref{fig:Fig_zy1}). Macroscale or partscale approaches intend to predict physical fields such as temperature, residual stresses, and thermal distortion on the scale of design parts~\citep{Gusarov2009, Zaeh2010, Denlinger2015, Hodge2016, Riedlbauer2017, Roy2018, Bartel2018,Zhang2018a, Kollmannsberger2019, Neiva2019, hodge_towards_2020, Proell2021}. Typically, the powder and melt phase are described in a spatially homogenized and simplified sense, e.g., without resolving the geometry and dynamics of powder particles and fluid flow in the melt pool. In contrast, mesoscale approaches resolve the length scale of individual powder particles. Usually, domains smaller than one powder layer are considered to either study the melt pool thermo-fluid dynamics during melting~\citep{Korner2013, Otto2012, Tan2014, Lee2015, Qiu2015, Khairallah2016, Russell2018, Yan2018,Weirather2019, Kouraytem2019, furstenau2020generating, Meier2020} or the cohesive powder dynamics during the previous powder spreading process~\citep{Herbold2015, Mindt2016, Haeri2017, Meier2019_2, Lee2020, HAN2019, wang2020, nan2020a, he2020, KOVALEV2020, shaheen2020, desai2019}. The former category of models aims at the prediction of melt pool instabilities (e.g., Rayleigh-Plateau or keyhole instabilities) and associated part defects such as residual porosity, lack of fusion, pure surface finish and insufficient layer-to-layer adhesion. The latter category of models intend to correlate parameters of the powder feedstock and the spreading process with resulting powder layer characteristics such as packing density and surface uniformity, and to understand the underlying powder-rheological mechanisms. Last, microscale approaches aim at predicting microstructure evolution during solidification and subsequent solid-state transformations in terms of phase composition, grain size, morphology (shape) and orientation, either based on spatially resolved grain/crystal geometries~\citep{Malinov.2000, Mishra.2004, Yang.2000, Ding.2004, Grujicic.2001, Kar.2006, reddy2008prediction, koepf2019numerical, nie2014numerical, Katzarov.2002, Radhakrishnan.2016, Chen.2004, gong2015phase} or on a spatially homogenized continuum representation~\citep{Murgau.C, Murgau.D, Ronda.1996, Malinov.2001.Differential, Luetjering.1998, Fan.2005, Crespo.2011, Porter.2009, Furrer.2010, Grong.2002, salsi2018modeling, lindgren2016simulation}.

The present article gives an overview of recent research activities and developments in these fields by our research group at the Institute for Computational Mechanics of the Technical University of Munich (TUM) together with our collaborators from the Massachusetts Institute of Technology (MIT), Lawrence Livermore National Laboratory (LLNL), the University of Innsbruck, and the Hamburg University of Technology. Beyond contributions to the individual disciplines of partscale thermo-mechanical modeling ({Section~\ref{sec:macro_modeling}}), mesoscale melt pool ({Section~\ref{sec:melt_pool_modeling}}) and powder modeling ({Section~\ref{sec:powder}}), and microstructure modeling ({Section~\ref{sec:micro_modeling}}), the question will be addressed how information between these different models can be exchanged and individual insights can be combined to achieve a holistic picture of the entire process chain in metal PBFAM ({Section~\ref{sec:wholistic}}). All models have been implemented in our in-house parallel multi-physics research code BACI~\cite{baci} jointly developed at the Institute for Computational Mechanics. Also the corresponding simulation results have been produced with BACI unless stated otherwise.

\section{Mesoscale Powder Modeling} \label{sec:powder}

\noindent

The characteristics of feedstock powders, and the resulting spreading kinematics critically influence the packing density and quality of the powder layer in PBFAM, which in turn is coupled to the melting process~\cite{Vock2019, Ziaee2019}.  In the authors' recent contribution~\cite{Meier2019}, a model for cohesive metal powders has been proposed and was subsequently applied to study the powder spreading process in metal PBFAM~\cite{Meier2019_2,Penny2020}. Important model equations and exemplary results are recapitulated in the following.

\subsection{Model equations} \label{sec:powder_equations}

\noindent
The proposed model is based on the discrete element method (DEM)~\cite{Cundall1979,Bashira1991,Bolintineanu2014,Thornton2000} and represents bulk powder mechanics on the level of individual powder particles in a Lagrangian manner. This model describes the kinematics of powder particles in 3D space by six degrees of freedom, i.e., the position vector $\mb{r}_G$ of the particle centroid as well as the rotation vector $\boldsymbol{\psi}$ and associated angular velocity $\boldsymbol{\omega}$. {In the present work, the powder particles are assumed to be spherical, which is a good approximation for the considered plasma-atomized powders, and follow a log-normal type size distribution~\cite{Meier2019}. Of course, a DEM representation of more general particle shapes, e.g., based on multi-sphere approaches, is possible as well.} The equations of motion of a particle $i$ follow from the balance of linear and angular momentum:
\begin{subequations}
\label{momentum}
\begin{align}
(m \, \ddot{\mb{r}}_G)^i = m^i \mb{g} + \sum_j (\mb{f}_{CN}^{ij}+\mb{f}_{CT}^{ij}+\mb{f}_{AN}^{ij}),\label{gusarov2007_HCE1}\\
(I_G \, \dot{\boldsymbol{\omega}})^i = \sum_j (\mb{m}_{R}^{ij}+\mb{r}_{CG}^{ij} \times \mb{f}_{CT}^{ij}).\label{gusarov2007_HCE2}
\end{align}
\end{subequations}
Here, $m\!=\!4/3\pi r^3\rho$ is the particle mass, $I_G=0.4mr^2$ is the moment of inertia of mass with respect to the particle centroid $G$, $r$ is the particle radius, $\rho$ is the mass density, and $\mb{g}$ is the gravitational acceleration. The interaction forces and torques between particles $i$ and $j$ on the right-hand side of~\eqref{momentum} consist of normal contact forces $\mb{f}_{CN}^{ij}$, tangential (frictional) contact forces $\mb{f}_{CT}^{ij}$, adhesive forces $\mb{f}_{AN}^{ij}$, and rolling resistance torques $\mb{m}_{R}^{ij}$. Here, $\mb{r}_{CG}^{ij}:=\mb{r}_{C}^{ij}-\mb{r}_{G}^i$ is the vector from the center of particle $i$ to the contact point with particle $j$. As shown in~\cite{Meier2019} and the subsequent work~\cite{Meier2019_2}, adhesive forces, mainly resulting from short range van der Waals (vdW) interactions between particles, dominate the dynamics of micron-scale metal powders and critically drive their spreading behavior during the powder recoating process in PBFAM. In~\cite{Meier2019}, an inter-particle force law, following the Derjaguin-Muller-Toporov (DMT) model, has been proposed:
\begin{align}
\label{adhesion5}
     F_S(s)=\left\{\begin{array}{ll}
      								F_{S0}=-4\pi \gamma r_{eff}, & g_N \leq  g_0\\
                                 \frac{A r_{eff}}{6s^2} , & g_0 < g_N < g^*\\
                                 0, & g_N \geq  g^*
                    \end{array}\right. 
                    \quad \text{with} \,\,
   g_0 := \sqrt{\frac{A r_{eff}}{6 F_{S0}}}, \quad g^* := \sqrt{\frac{1}{c_{FS0}}\frac{A r_{eff}}{6 F_{S0}}} = \frac{g_0}{\sqrt{c_{FS0}}}.
\end{align}
Based on~\eqref{adhesion5}, the adhesive forces read $\mb{f}_{AN}^{ij}=F_S(s)\mb{n}$, with the normal vector $\mb{n}=(\mb{r}_{G}^j-\mb{r}_{G}^i)/||\mb{r}_{G}^j-\mb{r}_{G}^i||$. Moreover, $\gamma$ is the surface energy, $A$ is the Hamaker constant, $r_{eff}:=\frac{r_i r_j}{r_i + r_j}$ is the effective interaction radius, $g_0$ is the distance at which the vdW force equals the pull-off force $F_{S0}$, and $g^*$ is defined as the cut-off radius at which the vdW has a relative decline of $c_{FS0}:=F_S(g^*)/F_{S0}$ with respect to the pull-off force $F_{S0}$ (taken as $c_{FS0}=1\%$ in our studies). In~\cite{Meier2019}, the surface energy $\gamma$ has been identified as most critical model parameter because: (i) it might vary by several orders of magnitude for a given material due to variations in particle surface roughness/topology and chemistry, and therefore cannot be taken from standard data bases but rather has to be identified/calibrated for the specific powder material; and (ii) it uniquely defines the magnitude of the particle pull-off force $F_{S0}$, which is critical for the flowability of these powders. For effective surface energy values in the range of $\gamma=0.1mJ/m^2$ as identified for these powders~\cite{Meier2019} and mean particle diameters in the range of $d=2r \approx 30 \mu m$, the adhesive forces according to~\eqref{adhesion5} are already by one order of magnitude higher than the gravity forces acting on these particles.

\subsection{Exemplary simulation results} \label{sec:powder_results}
\subsubsection{Powder Rheology} \label{sec:melt_powder_rheology}

\begin{figure}[htbp]
\centering
\subfigure [Simulated (left) vs. measured (right) static AOR~\cite{Meier2019}: coarse (top), medium (middle) and fine (bottom) powder.]
{
\includegraphics[height=0.43\textwidth]{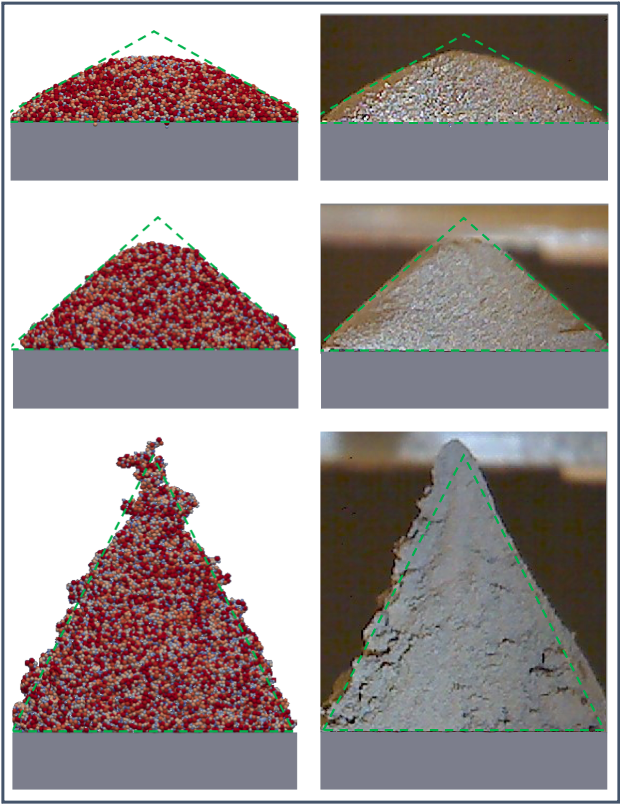}%
\label{fig:rheology1a}
}
\hspace{-0.00\textwidth}
\subfigure [Simulated (top) vs. measured (bottom) dynamic AOR in rotating drum.]
{
\includegraphics[height=0.43\textwidth]{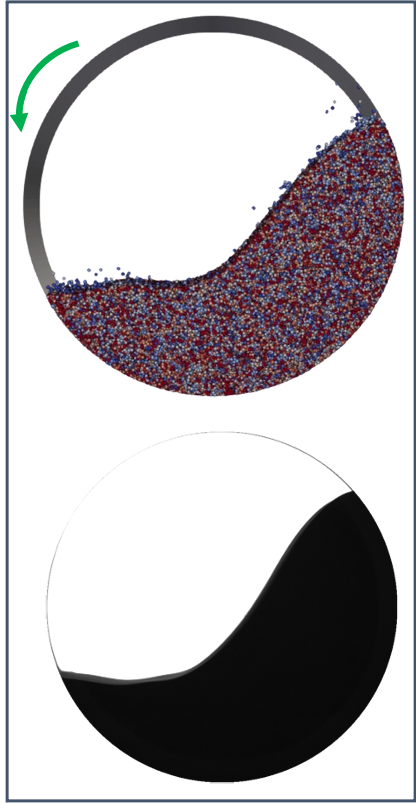}%
\label{fig:rheology1b}
}
\hspace{-0.00\textwidth}
\subfigure [Simulation-based (left) vs. experimental (right) setup for powder rheometry test based on rotating vane.]
{
\includegraphics[height=0.43\textwidth]{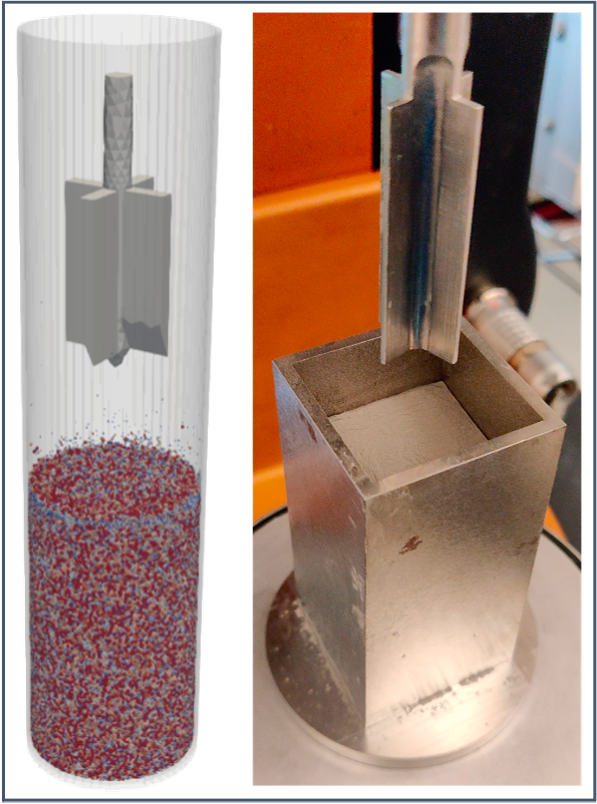}%
\label{fig:rheology1c}
}
\caption{Experimental and simulation-based realization of different static and dynamic powder rheology tests.}
\label{fig:rheology1}
\end{figure}

\noindent
Figure~\ref{fig:rheology1} represents different powder rheological setups along with their experimental and model-based realization. Specifically, funnel tests according to 
Figure~\ref{fig:rheology1a} allow for \textit{static} angle of repose (AOR) measurements, rotating drum tests according to 
Figure~\ref{fig:rheology1b} allow for \textit{dynamic} angle of repose (AOR) measurements, and rotating-vane rheometers allow to measure reaction torque curves as function of angular velocity $\Omega$ and compressing normal force $F_N$, which both can be controlled in these experiments. 
The purpose of such rheometer studies in the context of metal AM powder spreading process is two-fold: (i) to identify, i.e., calibrate, unknown parameters underlying the DEM model, which is essential to capture the physical powder behavior with sufficient accuracy; and (ii) to determine if the spreadability of a given powder material can be directly predicted by measurement of key rheological parameters such as static/dynamic AOR or powder characteristics gained from the rotating-vane rheometry. As emphasized in the last section, specifically the precise calibration of the effective surface energy $\gamma$ associated with the given powder material is of highest importance to achieve an accurate and predictive powder model. {The simulation results presented in the subsequent section are based on a calibrated model, whose effective surface energy has been identified from experimental static AOR measurements resulting in a value of $\gamma=0.1mJ/m^2$ ~\cite{Meier2019}}.

\subsubsection{Powder Spreading} \label{sec:melt_powder_spreading}

\begin{figure}[b!!!]
\centering
\subfigure [Rigid blade.]
{
\includegraphics[height=0.15\textwidth]{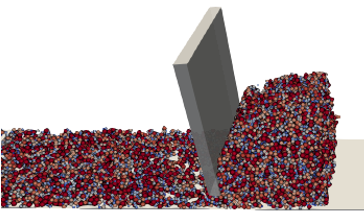}%
\label{fig:spreading1a}
}
\hspace{-0.01\textwidth}
\subfigure [Flexible blade.]
{
\includegraphics[height=0.15\textwidth]{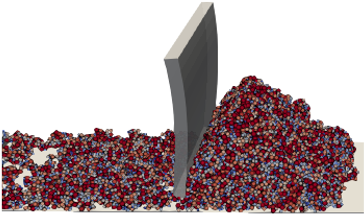}%
\label{fig:spreading1b}
}
\hspace{-0.01\textwidth}
\subfigure [Rotating roller.]
{
\includegraphics[height=0.15\textwidth]{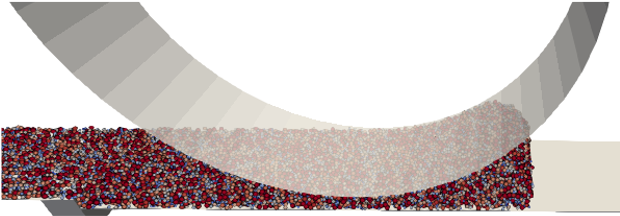}%
\label{fig:spreading1c}
}
\caption{Spreading simulations with different spreading tools: rigid blade, flexible blade and rotating roller.}
\label{fig:spreading1}
\end{figure}

\begin{figure}[b!!!]
   \centering
   \subfigure[{Influence of powder cohesiveness and nominal layer thickness on resulting mean value of spatial packing fraction field $\Phi(x,y)$~\cite{Meier2019_2}.}]
    {
    \includegraphics[width=0.40\textwidth]{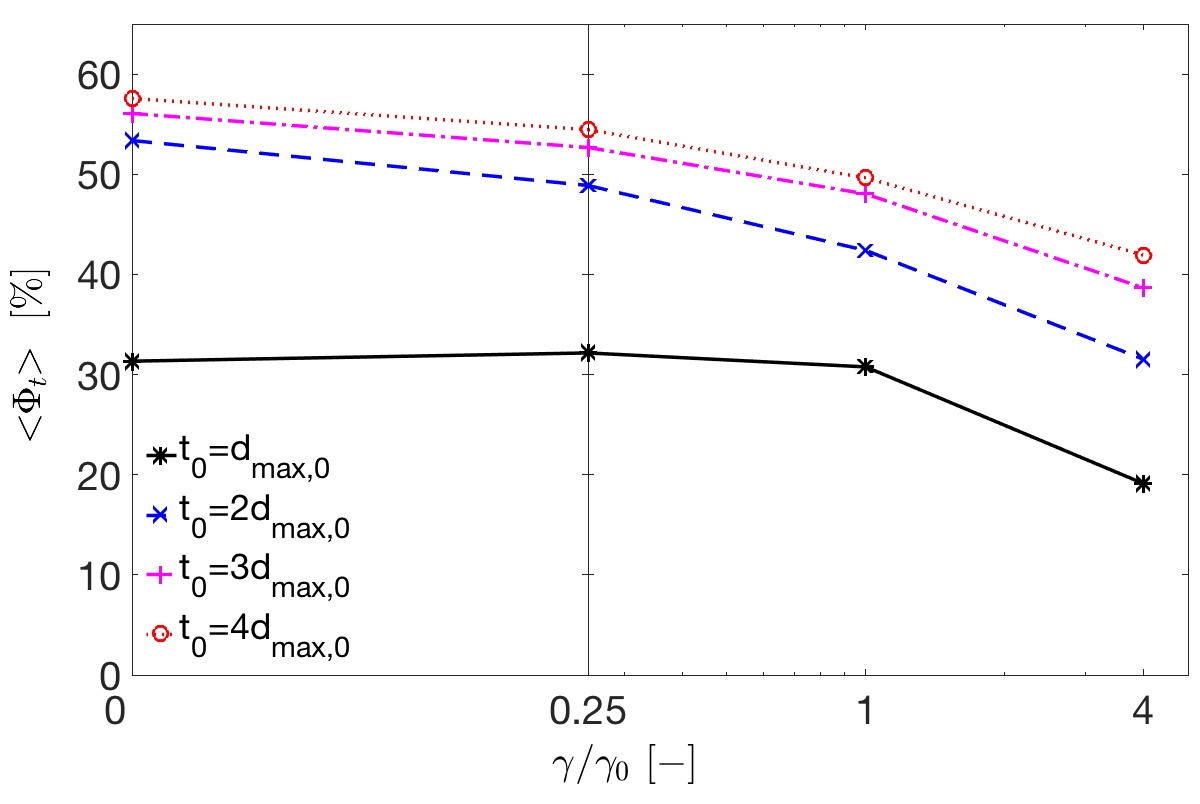}
    \label{fig:thickness_1}
   }
  \hspace{0.1 cm}
   \centering
 \subfigure[{Experimental measurement and associatd computational model prediction for mean value of spatial packing fraction field $\Phi(x,y)$~\cite{Penny2020}.}]
   {
    \includegraphics[width=0.40\textwidth]{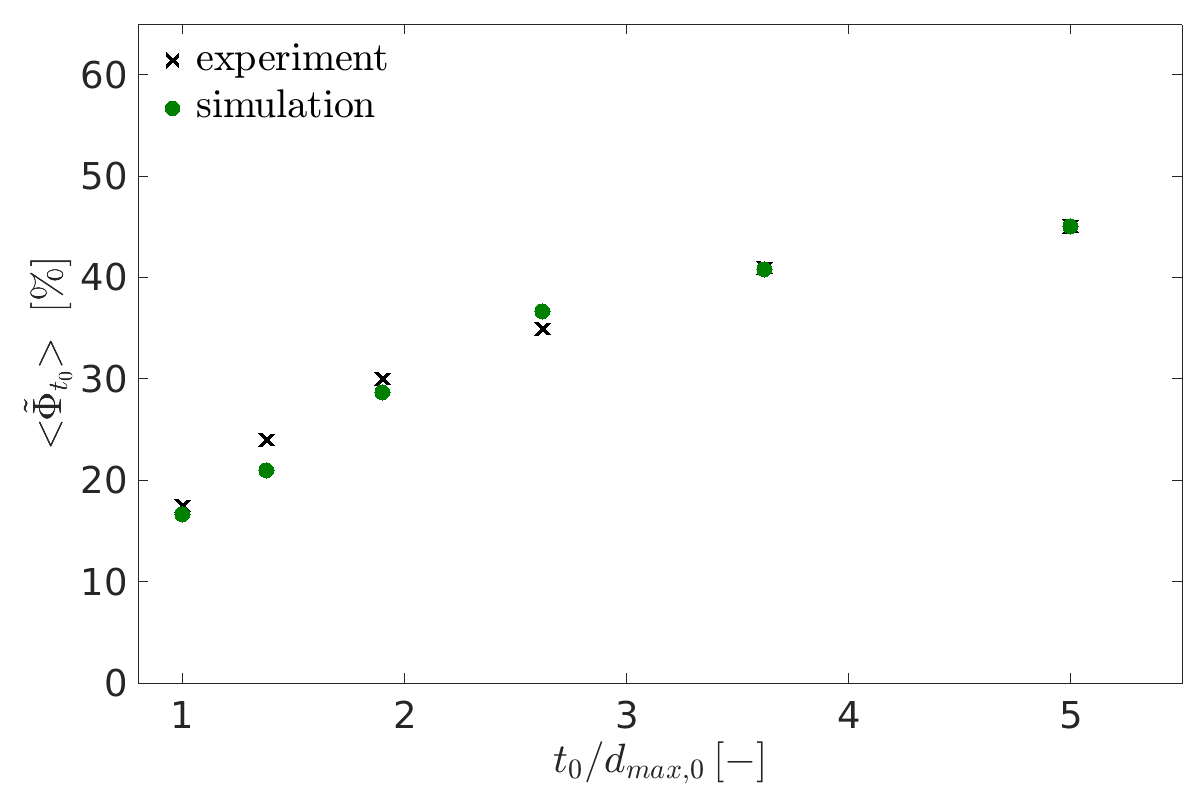}
    \label{fig:thickness_2}
   }    
   \caption{Mean value of spatial packing fraction field $\Phi(x,y)$ for different powders and layer thicknesses.}
  \label{fig:thickness}
\end{figure}

\begin{figure}[b!!!]
   \centering
    \subfigure[Standard deviation of packing fraction~\cite{Meier2019_2}.]
    {
    \includegraphics[width=0.40\textwidth]{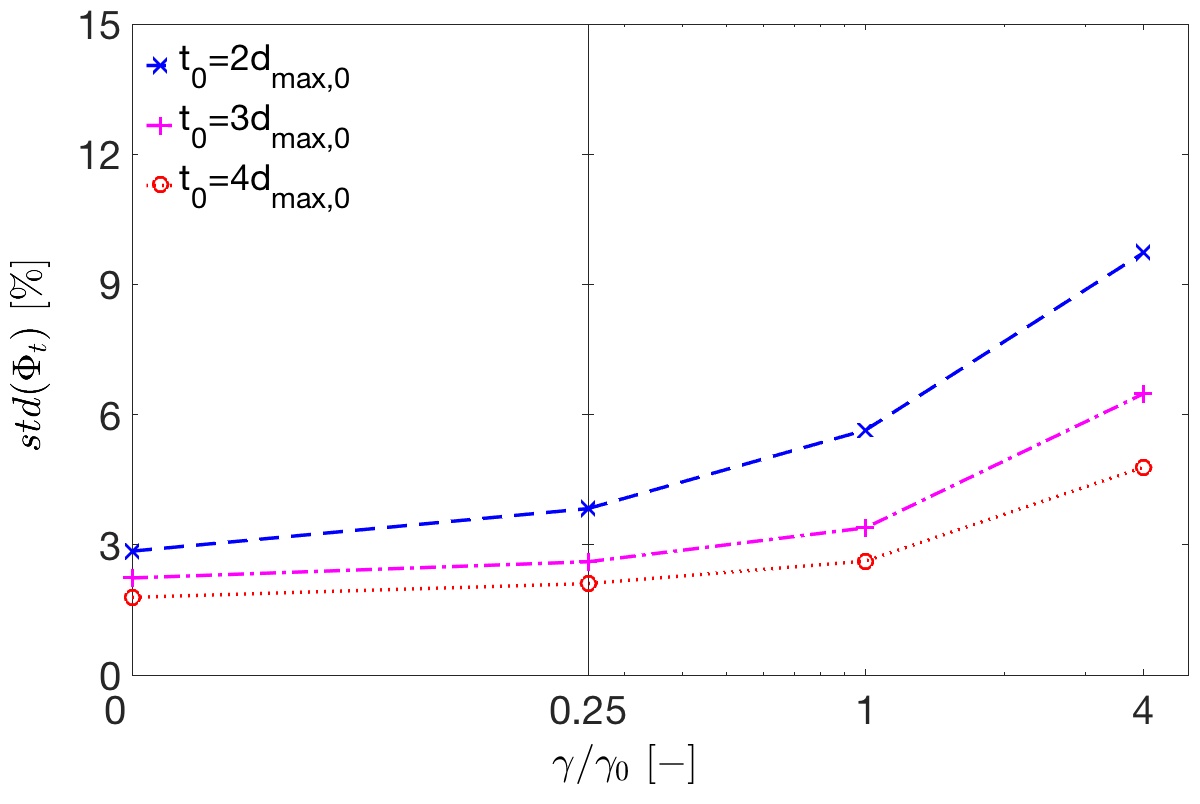}
    \label{fig:thickness_3}
   }
   \hspace{0.1 cm}
 \centering
 \subfigure[Standard deviation of surface profile~\cite{Meier2019_2}.]
   {
    \includegraphics[width=0.40\textwidth]{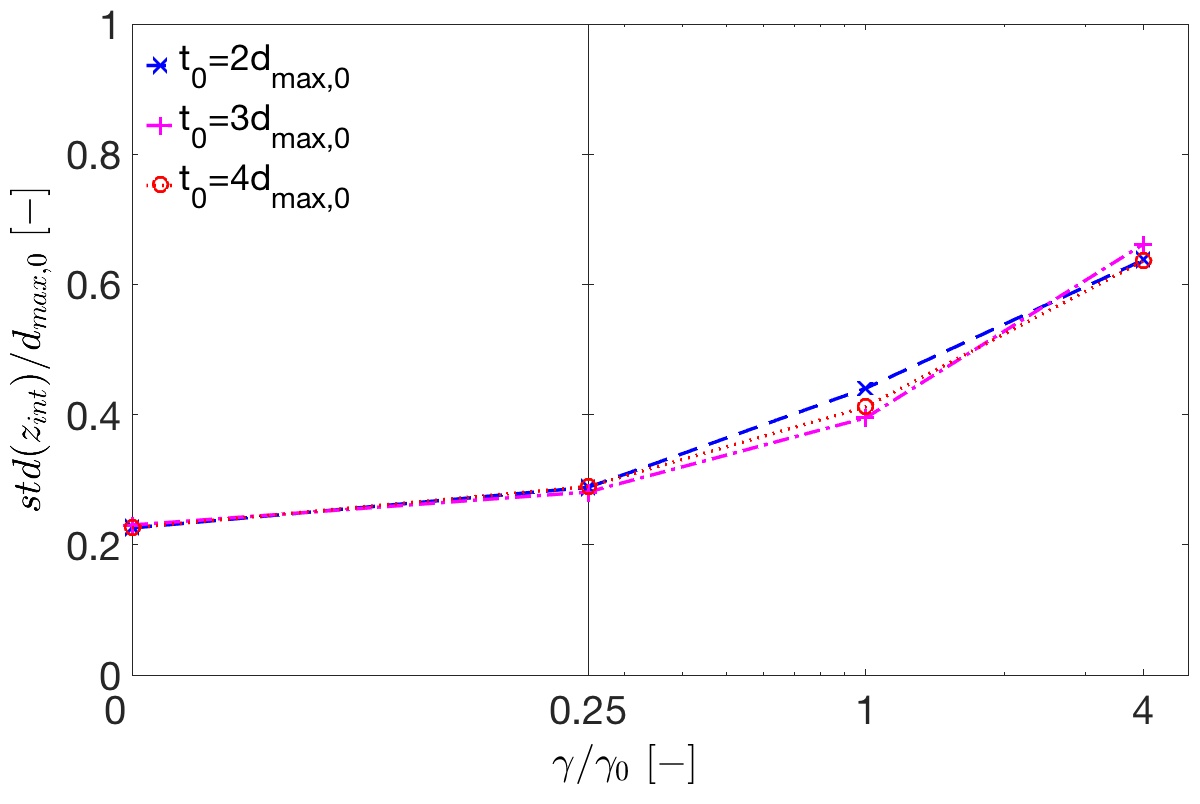}
    \label{fig:thickness_4}
   }
   \caption{Standard deviation of spatial packing fraction field $\Phi(x,y)$ and spatial surface profile $z_{int}(x,y)$.}
  \label{fig:thickness}
\end{figure}

\noindent
In~\cite{Meier2019_2}, a powder model as described above has been employed to study the powder spreading process in metal PBFAM using a rigid blade as recoating tool (see Figure~\ref{fig:spreading1a}). The mean values $<\!\! ... \!\!>$ as well as the standard deviation $std(...)$ of the spatial packing fraction field $\Phi(x,y)$ and the spatial surface profile $z_{int}(x,y)$ have been defined as metrics {(see~\cite{Meier2019_2} for the exact definition)} to rigorously assess powder layer quality. Special focus was on the influence of powder cohesiveness, representing powders with different mean particle size, on the resulting layer characteristics. Typical results are shown in Figures~\ref{fig:thickness_1},~\ref{fig:thickness_3} and~\ref{fig:thickness_4}. Accordingly, the powder layer quality decreases with increasing cohesiveness, which becomes visible through decreasing mean values of the packing fraction field and increasing standard deviations of packing fraction and surface profile, with the latter being a metric for surface roughness.
In addition, the influence of the nominal layer thickness $t_0$ as multiple of the (theoretical) maximal particle diameter $d_{max,0}$ (e.g., taken as $d_{max,0}=50\mu m$ for a powder size distribution with 90th percentile $D90=44\mu m$) is illustrated in these figures. For example, the mean packing fraction considerably increases with increasing nominal layer thickness approaching a saturation value at approximately $t_0/d_{max,0} \approx 4$. Figure~\ref{fig:thickness_2} shows recent results, where packing fraction values from spreading simulations and experimental spreading studies are compared, employing a novel X-ray microscopy technique for packing fraction measurement.
Accordingly, experiments and simulations are in very good agreement, confirming that packing fraction increases with increasing nominal layer thickness $t_0$ and approaching a saturation value in the range of $t_0/d_{max,0} \approx 4$. Note that slightly different definitions of the reference volume underlying the packing fraction calculations in Figures~\ref{fig:thickness_1} and~\ref{fig:thickness_2} have been applied. In addition to the rigid blade studies discussed so far, Figures~\ref{fig:spreading1b} and~\ref{fig:spreading1c} show results from recent spreading studies employing also flexible blades and rotating rollers. Specifically, the flexible blade simulations were enabled by a staggered coupling algorithm for structure-particle interaction~\cite{Fuchs2020} taking advantage of the finite element method (FEM) and discrete element method (DEM) modules in our research code BACI. In Table~\ref{tab:spreading_results_different_tools}, first results are depicted for the mean values and standard deviations of packing fraction and surface profile resulting from simulations with rigid and flexible blade respectively non-rotating and counter-rotating roller.

The simulations have been carried out for the most cohesive powder depicted in Figures~\ref{fig:thickness_1},~\ref{fig:thickness_3} and~\ref{fig:thickness_4} (i.e., $\gamma/\gamma_0=4$) with $d_{max,0}=50 \mu m$ and $t_0=3 d_{max,0}$. To predict an upper limit for the maximal achievable layer quality, particle-substrate adhesion has been set identical to particle-particle adhesion and particle-blade/roller adhesion has been set to zero~\cite{Meier2019_2}. As expected, the mechanical compression forces induced by the roller geometry lead to a considerably improved layer quality as compared to the rigid blade. Interestingly, the results from the non-rotating and counter-rotating roller are almost identical. This might be explained by the idealized wall-adhesion properties in this study. With non-zero particle-roller adhesion and potentially decreased particle-substrate adhesion, which might yield considerable stick-slip motions of the powder on the substrate~\cite{Meier2019_2}, a stronger dependence of the layer characteristics on the roller rotation is expected. Also the low layer quality resulting from the flexible blade is an unexpected result at first glance. However, from looking at the dynamic spreading motion it would become apparent that the powder-blade interaction leads to dynamic bending oscillations of the flexible blade, which explains the irregularity of the resulting powder layer. At this point, it is emphasized that the flexible blade setup considered here should rather serve as a proof of principle for the proposed modeling and simulation framework, while flexible blades applied in practice are typically of a geometrically more complex shape.

\begin{table}[h!!!]
\centering
\begin{tabular}{|p{4.0cm}|p{1.6cm}|p{1.8cm}|p{2.4cm}|p{3.2cm}|} \hline
 spreading tool & $<\!\Phi\!> \,\, [\%]$ & $std(\Phi) \,\,\,\, [\%]$ & $<\!z_{int}\!>\!/t_0 \,\,\,\, [-]$ & $std(z_{int})/d_{max,0} \,\,\,\, [-]$ \\ \hline
rigid blade & $39.6$ & $8.8$ & $0.64$ & $0.79$ \\ \hline
flexible blade & $32.1$ & $11.7$ & $0.54$ & $1.14$ \\ \hline
non-rotating roller & $50.5$ & $2.9$ & $0.99$ & $0.20$ \\ \hline
counter-rotating roller & $51.3$ & $3.1$ & $0.98$ & $0.20$ \\ \hline
\end{tabular}
\caption{Powder layer metrics resulting from simulations with different spreading tools.}
\label{tab:spreading_results_different_tools}
\end{table}

\subsection{Experimental validation}
\label{powder_exp}
{Figure~\ref{fig:thickness_2} demonstrates experimental validation of the DEM powder simulations using an X-ray microscopy technique recently developed at MIT~\cite{Penny2020}. This method was developed to provide direct interrogation of local powder deposition, in contrast with optical interrogation techniques (e.g., \cite{Craeghs2011_2, Kleszczynski2012, Hendriks2019, Snow2019, TanPhuc2019AManufacturing}) that accurately assess layer topography but cannot reliably determine the volume of material deposited as packing density is also spatially varying~\cite{Ali2018OnProcesses, TanPhuc2019AManufacturing}.  Herein, etched silicon templates, manufactured to sub-micron tolerances of nominal depth, flatness, and surface roughness, provide precision control of powder geometry.  Specimen powder layers are created by sweeping powder into the template, and any variation in powder deposition may be fully ascribed to stochastic powder flow and not to disturbances from poorly controlled boundary conditions.  X-ray transmission of the powder layer is assessed by first imaging the empty template and again after creating a layer specimen; these data are interpreted as an effective thickness of metal powder using a radiation transport model considering: the polychromatic X-ray source emission spectrum, wavelength-dependent absorption spectrum of objects (e.g., template and powder layer) in the beam path, and detector physics including scintillation and signal gain.}

{Using this technique, experimental conditions may be matched as closely as possible to simulations for validation. The aforementioned figure demonstrated close agreement in which commercially obtained 15-45 $\mu m$ Ti-6Al-4V powder has been spread into layers nominally 50 to 250 $\mu m$ thick. After careful model calibration via static angle of repose as specified above, we observe that both the experimental measurement and model prediction of packing fraction asymptotically increase with increasing layer thickness as particles are able to assume more dense configurations as compared to cases where layer thickness approaches $d_{max}$ (particle size).  More broadly, the experiment reproduces key relationships between powder flowability, where high angle of repose powders create comparatively sparse layers with high variance in deposition.}

\section{Mesoscale Modeling of Melt Pool Thermo-Hydrodynamics} \label{sec:melt_pool_modeling}

\noindent

In the following, the most important model equations and exemplary results of a novel smoothed particle hydrodynamics (SPH) formulation for thermo-capillary phase change problems will be recapitulated, which has recently been proposed by the authors and applied to metal AM melt pool modeling~\cite{Meier2020}. 

\subsection{Model equations} \label{sec:melt_pool_modeling_sph_equations}

\noindent
The domain of melt pool thermo-hydrodynamics in metal AM is split into the liquid melt phase $\Omega_{l}$, the atmospheric gas phase $\Omega_{g}$ as well as the solid phase $\Omega_{s}$, allowing for reversible phase transition between liquid and solid phase. The liquid and gas phase are governed by the  \textit{weakly compressible}, anisothermal Navier-Stokes equations in the fluid domain~$\Omega^{f}=\Omega^{l} \cup \Omega^{g}$. The problem is described by a set of six equations, namely the continuity equation~\eqref{eq:SPH_conti}, the Navier-Stokes momentum equation~\eqref{eq:SPH_momentum} (3 components), the energy equation~\eqref{eq:SPH_energy} and an equation of state~\eqref{eq:SPH_eos} associated with the weakly compressible approach.
\begin{subequations}
\label{eq:SPH_basic}
\begin{align}
\dv{\rho}{t} = -\rho \div \vectorbold{u} \qin \Omega^{f},\label{eq:SPH_conti}\\
\dv{\vectorbold{u}}{t} = \frac{1}{\rho} \left(-\grad{p} + \vectorbold{f}_{\nu} + \tilde{\vectorbold{f}}^{{lg}}_{s} + \tilde{\vectorbold{f}}^{{slg}}_{w} + \tilde{\vectorbold{f}}^{{lg}}_{v} \right) + \vectorbold{g} \qin \Omega^{f},\label{eq:SPH_momentum}\\
c _p\dv{T}{t} =  \frac{1}{\rho} \left(\div  (k \grad{T}) + \tilde{s}^{lg}_v + \tilde{s}^{lg}_l \right)  \qin \Omega, \label{eq:SPH_energy}\\
p\qty(\rho) = c^{2} \qty(\rho - \rho_{0}) = p_{0} \qty(\frac{\rho}{\rho_{0}} - 1)  \qin \Omega^{f}. \label{eq:SPH_eos}
\end{align}
\end{subequations}
Note, that the energy equation is solved for the entire domain $\Omega = \Omega^{l} \cup \Omega^{g} \cup \Omega^{s}$, i.e., also for the solid phase.
The six primary unknowns are given by velocity $\vectorbold{u}$ (three components), density~$\rho$, pressure~$p$ and temperature~$T$. The momentum equation~\eqref{eq:SPH_momentum} contains contributions from viscous forces~$\vectorbold{f}_{\nu}= \eta \laplacian{\vectorbold{u}}$ with dynamic viscosity $\eta$, surface tension forces~$\tilde{\vectorbold{f}}^{{lg}}_{s}$, wetting forces~$\tilde{\vectorbold{f}}^{{slg}}_{w}$, evaporation-induced recoil pressure forces $\tilde{\vectorbold{f}}^{{lg}}_{v}$, and body forces~$\vectorbold{g}$. Here, the superscripts of interface forces refer to the corresponding interface, e.g., to the liquid-gas interface (lg) or the triple line solid-liquid-gas (slg). In particular, evaporation is modeled on basis of a phenomenological model for the evaporation-induced recoil pressure as proposed by Anisimov~\cite{Anisimov1995}:
\begin{equation} \label{eq:SPH_evap_recoil}
\tilde{\vectorbold{f}}^{{lg}}_{v} = -p_v(T) \vectorbold{n}^{lg} \delta^{lg} \quad \text{with} \quad p_v(T) = 0.54 p_a \exp \left[ - \frac{\bar{h}_v}{R} \left( \frac{1}{T} - \frac{1}{T_v} \right)\right],
\end{equation}
with the atmospheric pressure $p_a$, the molar latent heat of evaporation $\bar{h}_v$, the molar gas constant $R$, and the boiling temperature $T_v$. Here, $\vectorbold{n}^{lg}$ is the normal vector of the liquid-gas interface and $\delta^{lg}$ the corresponding surface delta function distributing interface surface forces across a diffuse interface domain of finite thickness. The energy equation~\eqref{eq:SPH_energy} contains the mass-specific heat capacity $c_p$, the heat flux $\vectorbold{q}:= -k \grad{T}$ according to Fourier's law (with thermal conductivity $k$) as well as heat fluxes stemming from the laser beam heat source $\tilde{s}^{lg}_l$ and from evaporation-induced heat losses $\tilde{s}^{lg}_v$. The former is modeled as a surface heat flux with Gaussian profile. The latter is consistent with the aforementioned recoil pressure model~\cite{Anisimov1995} and reads:
\begin{equation} \label{eq:SPH_evap_heatloss}
\tilde{s}^{lg}_v = s^{lg}_v \, \delta^{lg} \quad \text{with} \quad s^{lg}_v = - \dot{m}^{lg}_v [h_v+ h(T)], \quad \dot{m}^{lg}_v = 0.82 \, c_s \, p_v(T) \, \sqrt{\frac{M}{2\pi RT}}, \quad h(T)=\int \limits_{T_{h,0}}^T  c_p \, \, d\bar{T},
\end{equation}
where the enthalpy rate per unit area $s^{lg}_v$ results from the vapor mass flow per unit area $\dot{{m}}^{lg}_v$ at the melt pool surface and the sum of the specific enthalpy $h(T)$ and the latent heat of evaporation $h_v$, both per unit mass. Moreover, $T_{h,0}$ is a reference temperature of the specific enthalpy and $M$ is the molar mass. Finally, $p_v(T)$ is the recoil pressure defined in~\eqref{eq:SPH_evap_recoil} and $c_s$ the so-called sticking constant which takes on a value close to one, i.e., $c_s \approx 1$, for metals~\cite{Khairallah2016, Weirather2019}. Finally, in the equation of state~\eqref{eq:SPH_eos} the reference density~$\rho_{0}$, the reference pressure~$p_{0} = \rho_{0} c^{2}$ and the artificial speed of sound~$c$ can be identified. Note that the commonly applied weakly compressible approach only represents deviations from the reference pressure, i.e., $p\qty(\rho_{0}) = 0$, and not the total pressure. Equations~\eqref{eq:SPH_basic} are discretized in space by smoothed particle hydrodynamics (SPH) and in time by an explicit velocity-Verlet scheme. It is emphasized that SPH is a Lagrangian approach, i.e., material particles are used for discretization that are convected by the fluid velocity and directly carry the phase information without requiring additional phase tracking schemes. In particular, after spatial discretization the model equations~\eqref{eq:SPH_basic} (in strong form) are evaluated at the positions of these discretization particles taking into account the material parameters corresponding to the respective phase of a particle, i.e., solid, liquid or gas phase. Continuous primary variable fields are approximated by applying a smoothing kernel function to the discrete particle values of these variables, which is required to define spatial gradients.

\subsection{Exemplary simulation results} \label{sec:melt_pool_modeling_sph_results}

\begin{figure}[t!!!]
\centering
%
\subfigure [time: $t=0.079ms$]
{
\includegraphics[width=0.23\textwidth]{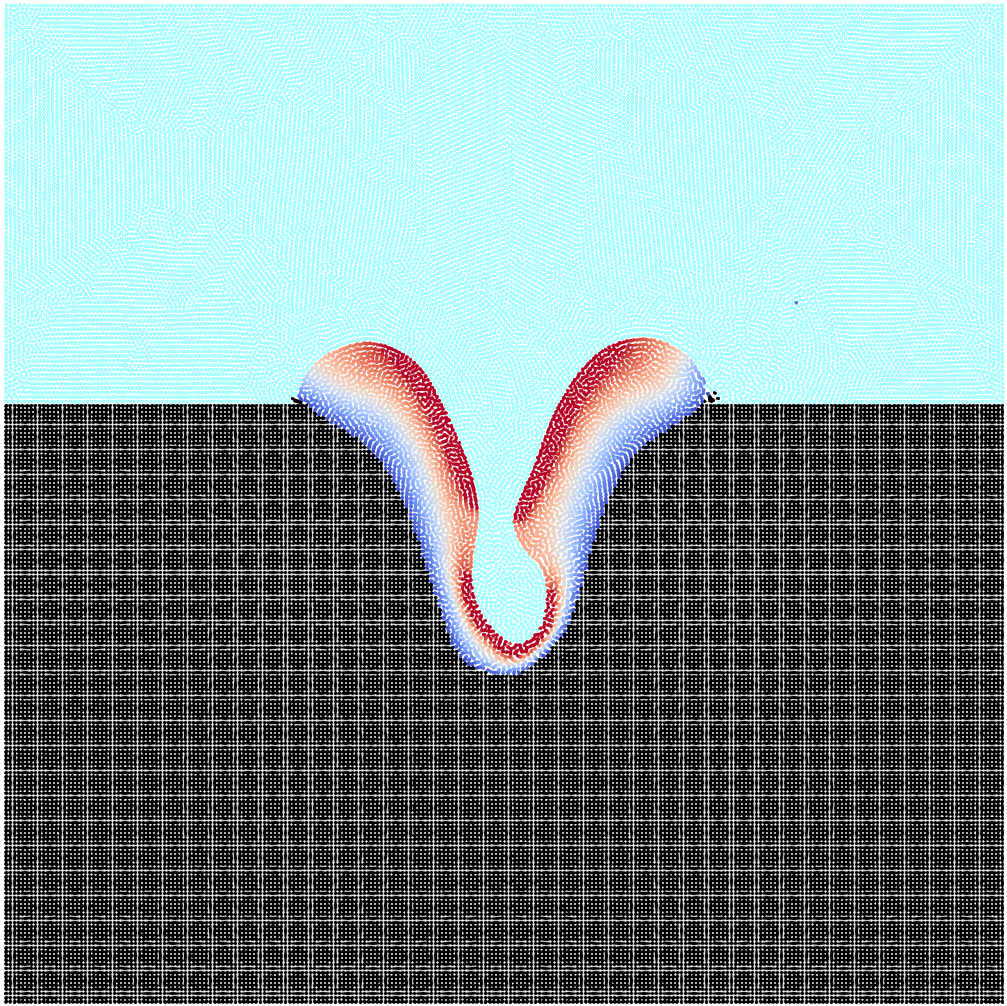}
\label{fig:example5_2_1}
}
\subfigure [time: $t=0.083ms$]
{
\includegraphics[width=0.23\textwidth]{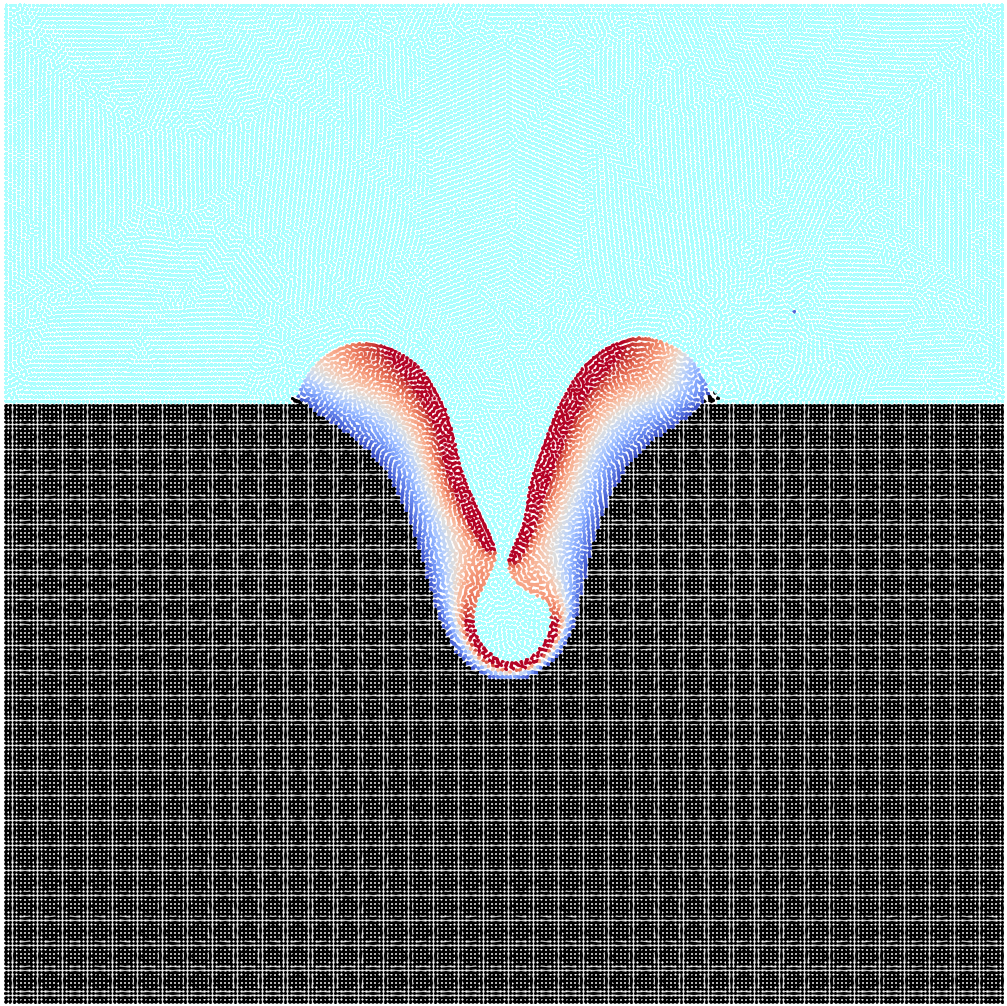}
\label{fig:example5_2_2}
}
\subfigure [time: $t=0.085ms$]
{
\includegraphics[width=0.23\textwidth]{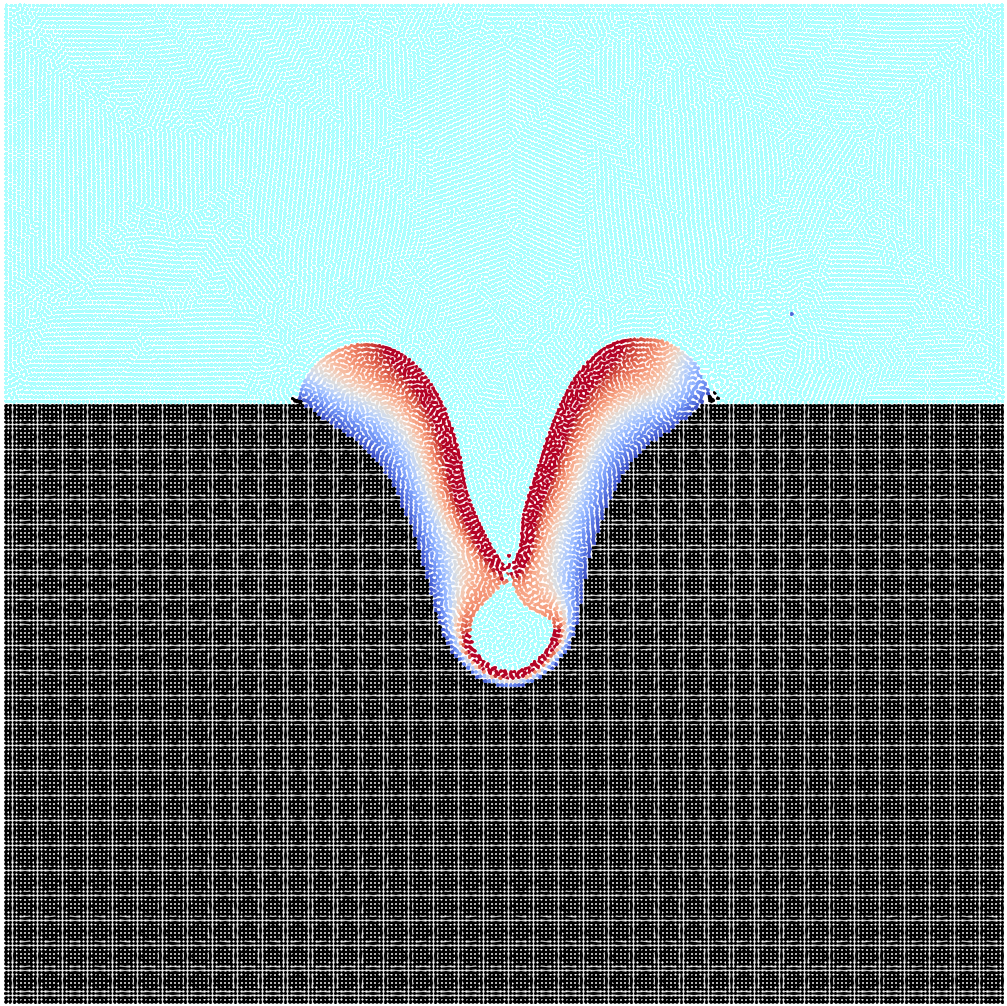}
\label{fig:example5_2_3}
}
\subfigure [time: $t=0.089ms$]
{
\includegraphics[width=0.23\textwidth]{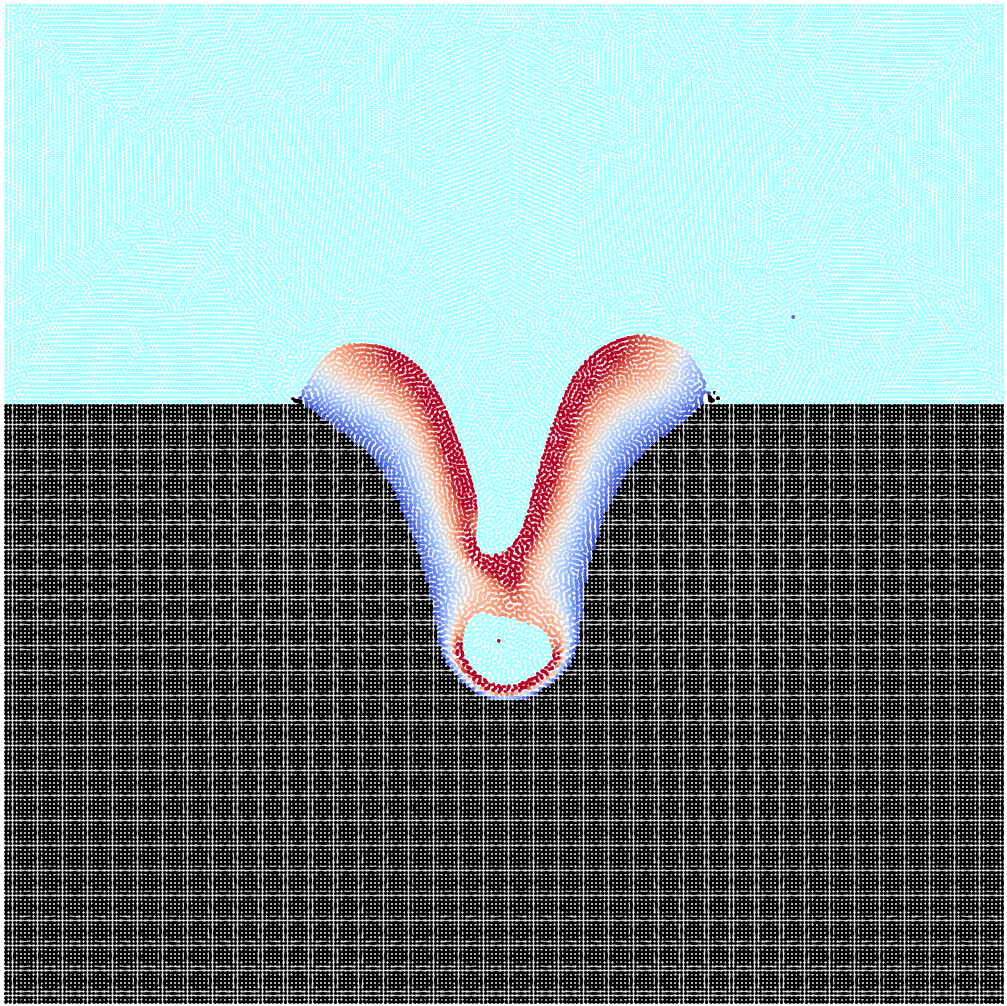}
\label{fig:example5_2_4}
}
\caption{2D laser melting with recoil pressure according to \textit{variant 1}: creation mechanism of a gas inclusion at different time steps. Temperature range from $1700K$ (blue) to $3500K$ (red). Display of solid phase in black and gas phase in light blue~\cite{Meier2020}.}
\label{fig:example5_2}
\end{figure}

\begin{figure}[t!!!]
\centering
%
\subfigure [time: $t=0.056ms$]
{
\includegraphics[width=0.23\textwidth]{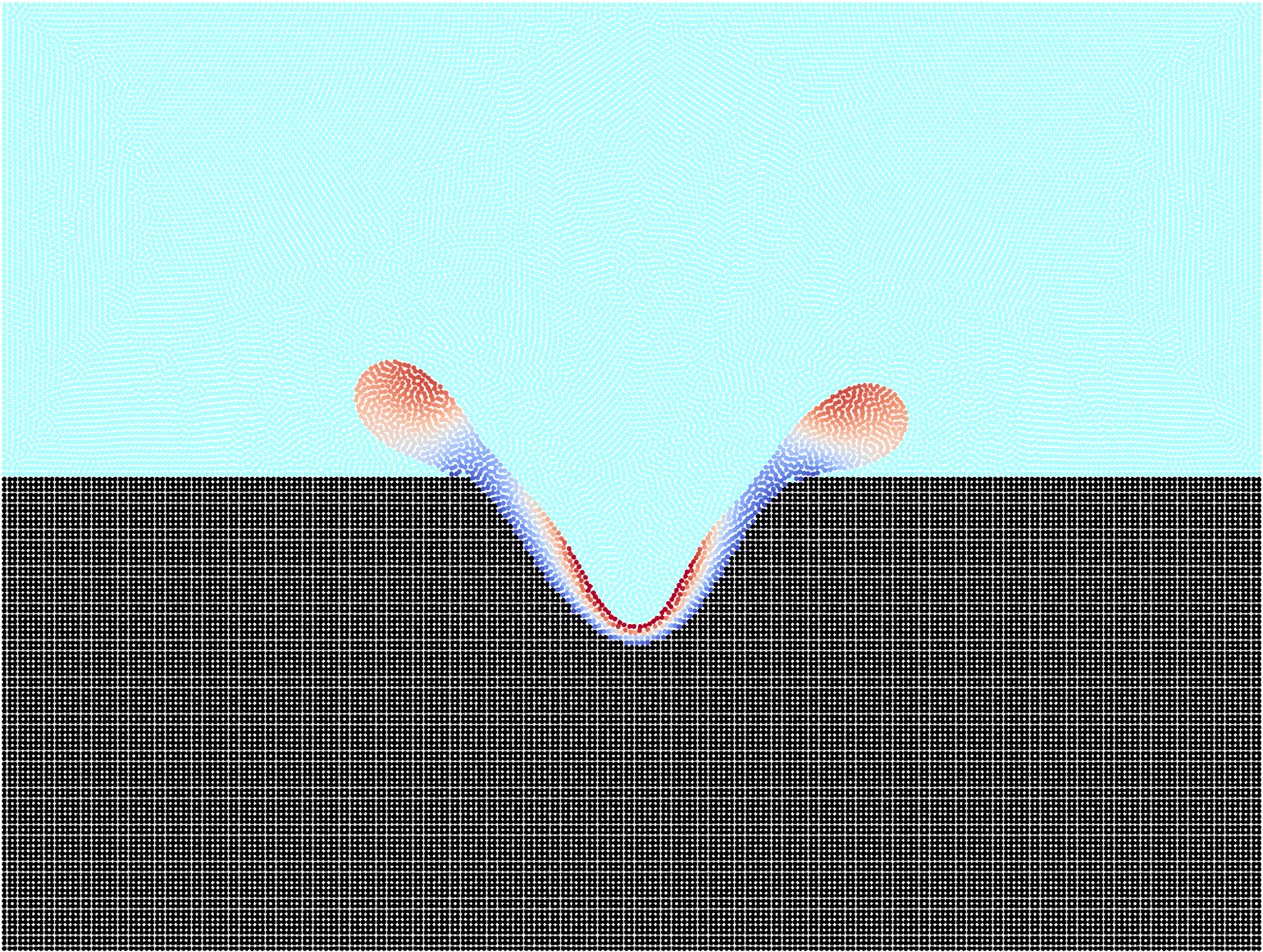}
\label{fig:example5_1_1}
}
\subfigure [time: $t=0.082ms$]
{
\includegraphics[width=0.23\textwidth]{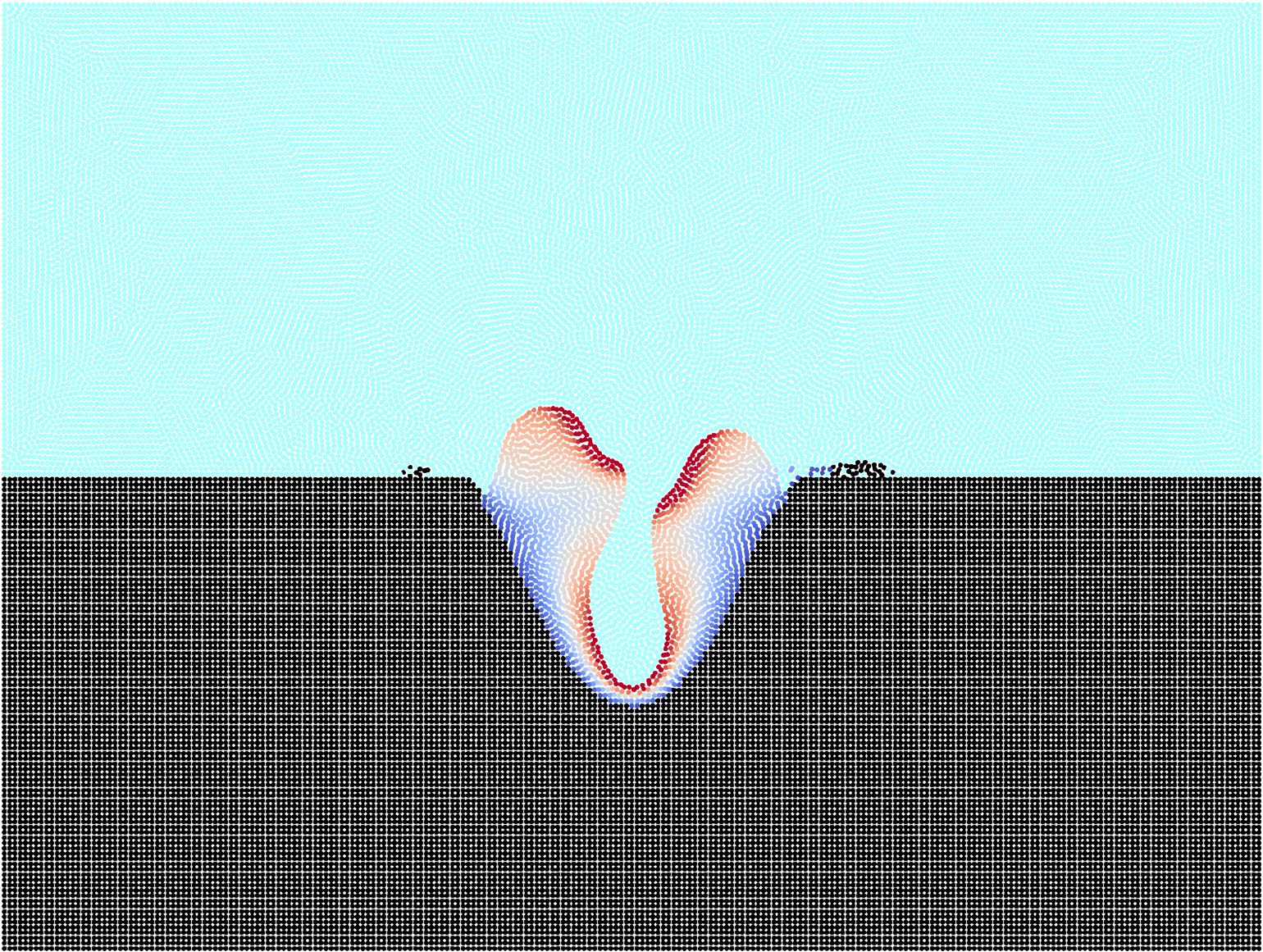}
\label{fig:example5_1_2}
}
\subfigure [time: $t=0.109ms$]
{
\includegraphics[width=0.23\textwidth]{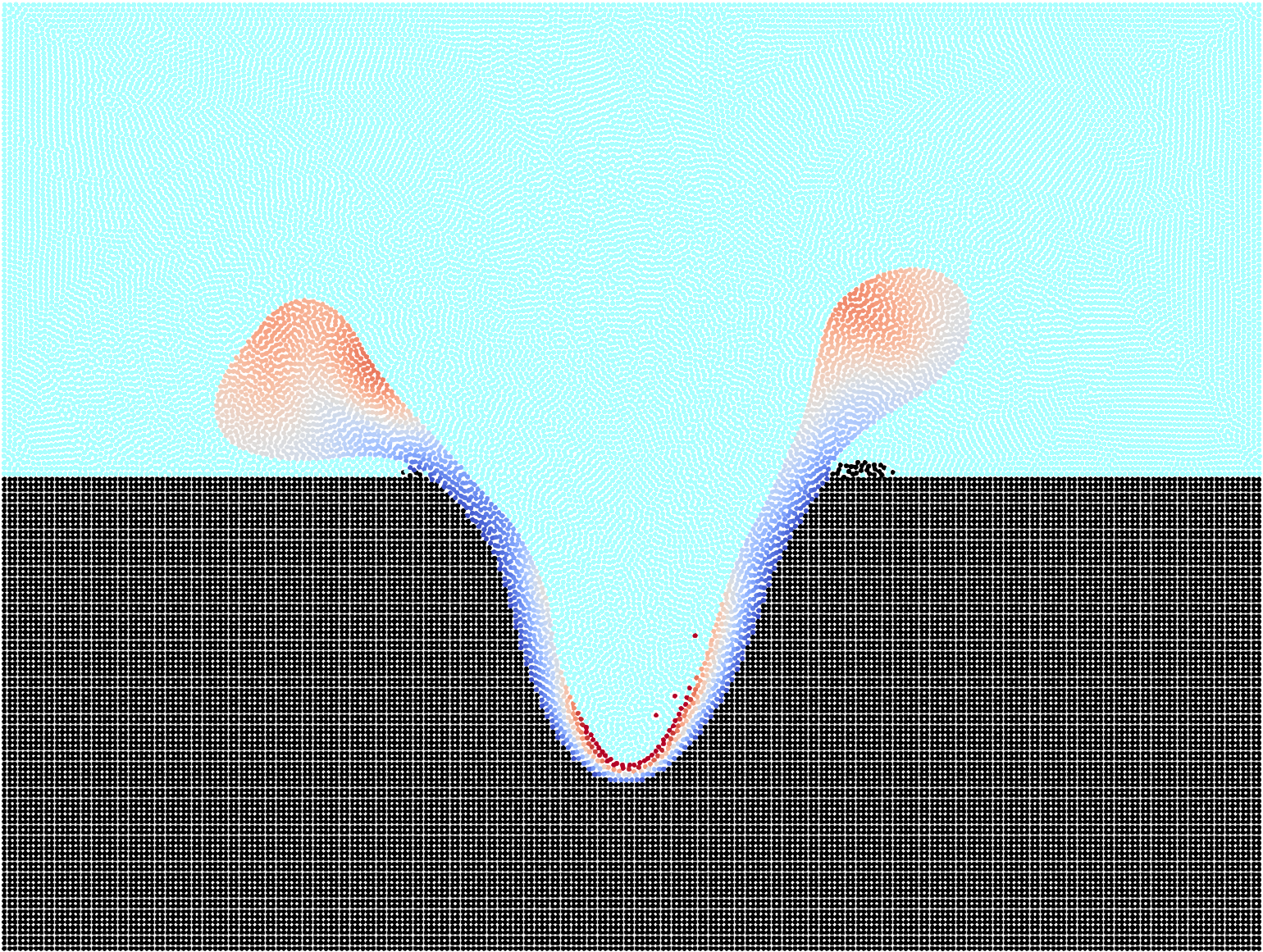}
\label{fig:example5_1_3}
}
\subfigure [time: $t=0.135ms$]
{
\includegraphics[width=0.23\textwidth]{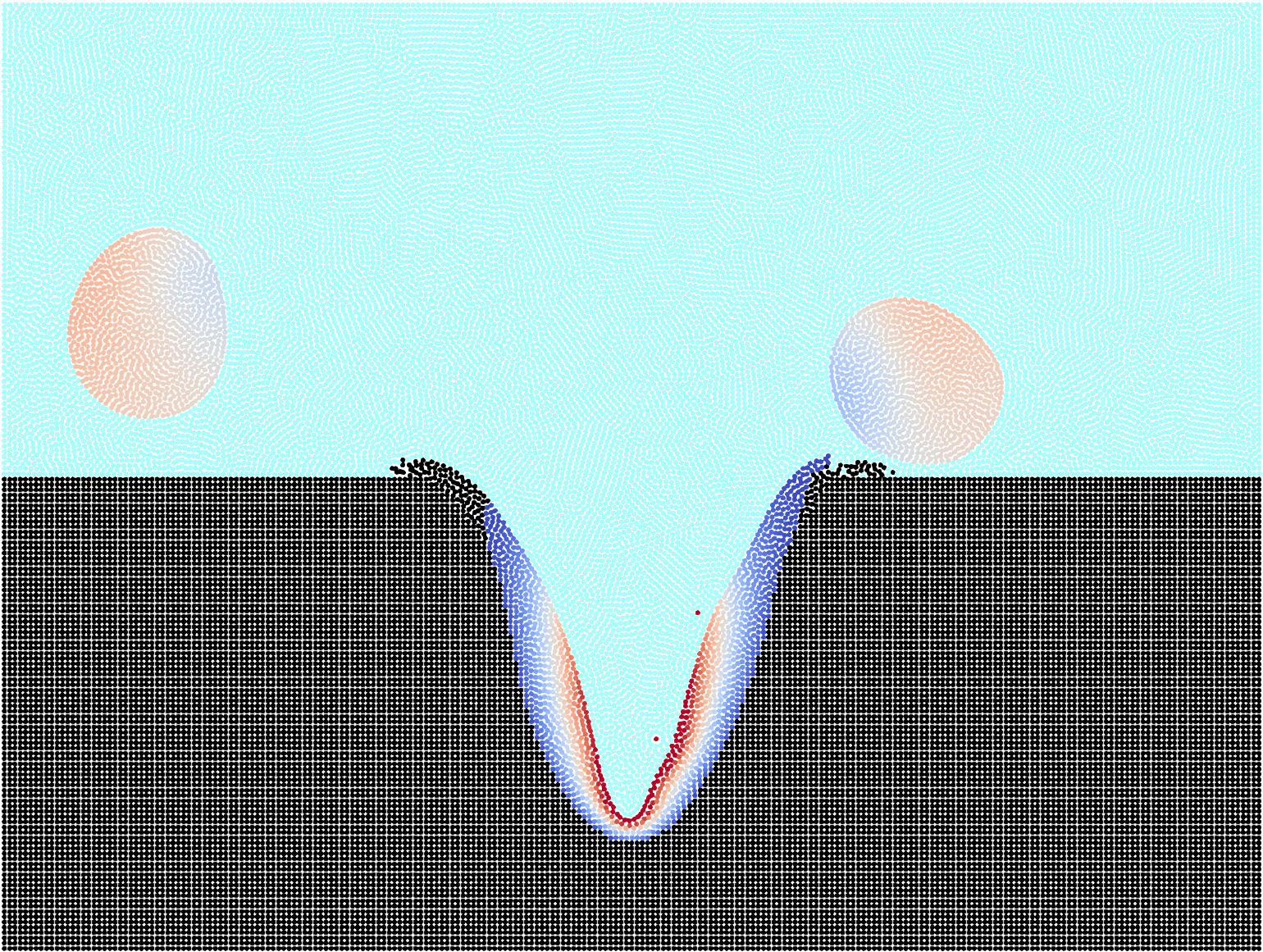}
\label{fig:example5_1_4}
}
\caption{2D laser melting with recoil pressure according to \textit{variant 2}: process of melt drop ejection at different time steps. Temperature range from $1700K$ (blue) to $3500K$ (red). Display of solid phase in black and gas phase in light blue~\cite{Meier2020}.}
\label{fig:example5_1}
\end{figure}

\noindent
While the accuracy of the individual model components has been critically verified and compared to analytical solutions in~\cite{Meier2020}, two examples with particular relevance to PBFAM melt pool modeling, will be recapitulated in the following. The results of two exemplary point melting examples, i.e., melting with a spatially fixed laser, in 2D are illustrated in Figures~\ref{fig:example5_2} and~\ref{fig:example5_1} and in the following denoted as \textit{variant 1} and \textit{variant 2}. The only difference between the two variants is that the evaporative heat loss term~\eqref{eq:SPH_evap_heatloss} has been neglected in the second variant according to Figure~\ref{fig:example5_1}. In order to end up with a comparable effective energy input into the system and a comparable level of peak temperatures in the melt pool, the laser power had to be reduced by a factor of $200$ for this second variant. On the one hand, this procedure demonstrates the importance of this heat loss term. On the other hand, it allows to study different physical phenomena and resulting melt pool characteristics, namely either melt droplets ejected from the melt pool (Figure~\ref{fig:example5_1}) or gas bubbles trapped in the melt pool (Figure~\ref{fig:example5_2}).
First, these two scenarios allow to verify the robustness of the proposed SPH formulation in representing highly dynamic changes of the liquid-gas interface topology involving the coalescence and separation of interface segments. On the one hand, these are process-typical scenarios when spatter or gas inclusions are created, e.g., in the keyhole regime of PBFAM. On the other hand, an accurate and robust description of such interface phenomena is typically rather challenging for mesh-free discretization schemes such as SPH, which underlines the robustness of the present formulation. Second, these two variants also allow to study the different physical phenomena underlying the mechanisms of spatter generation respectively gas bubble inclusion, thus gaining detailed understanding of process dynamics and instabilities in PBFAM. Figure~\ref{fig:example5_2} displays the system behavior of \textit{variant 1}. Accordingly, the interaction of surface tension and recoil pressure forces leads to oscillations of the liquid-gas interface with maximal amplitudes slightly above the bottom of the keyhole. Once, the amplitudes of these oscillations are large enough such that the opposite keyhole walls gets into touch a liquid bridge forms, which effectively encloses the gas material at the keyhole base - a gas inclusion is formed. Note again, that this example mainly aims at demonstrating the robustness of the proposed formulation in representing the highly dynamic surface tension-recoil pressure interaction and eventually the formation of a gas bubble. Of course, the employed phenomenological recoil pressure model does not explicitly resolve the high-velocity vapor jet that would arise from the keyhole base in the real physical system. This vapor jet might considerably influence the described creation mechanism of a liquid bridge, and potentially burst it before a closed gas bubble is created. In Section~\ref{sec:melt_pool_evaporation} first steps towards a high-fidelity melt pool model that explicitly resolves these effects of evaporation and gas dynamics will be presented. {Also first steps towards the representation of mobile powder particles in the melt pool model have been made~\cite{Fuchs2021}.}

Figure~\ref{fig:example5_1} displays the melt pool dynamics resulting from \textit{variant 2}. The high recoil pressure magnitudes in this variant result in periodic flow patterns consisting of recoil pressure-driven high-velocity waves from the center to the edges of the melt pool and surface tension-driven back flow from the edges to the center. The continued laser energy input results in increasing amplitudes of these flow oscillations until eventually melt droplets are ejected at the melt pool edges (Figures~\ref{fig:example5_1_3} and~\ref{fig:example5_1_4}).

\begin{figure}[h!!!!!!]
\centering
\subfigure [time: approx.  $0.12ms$ \hspace{0.22\textwidth}]
{
\includegraphics[width=0.48\textwidth]{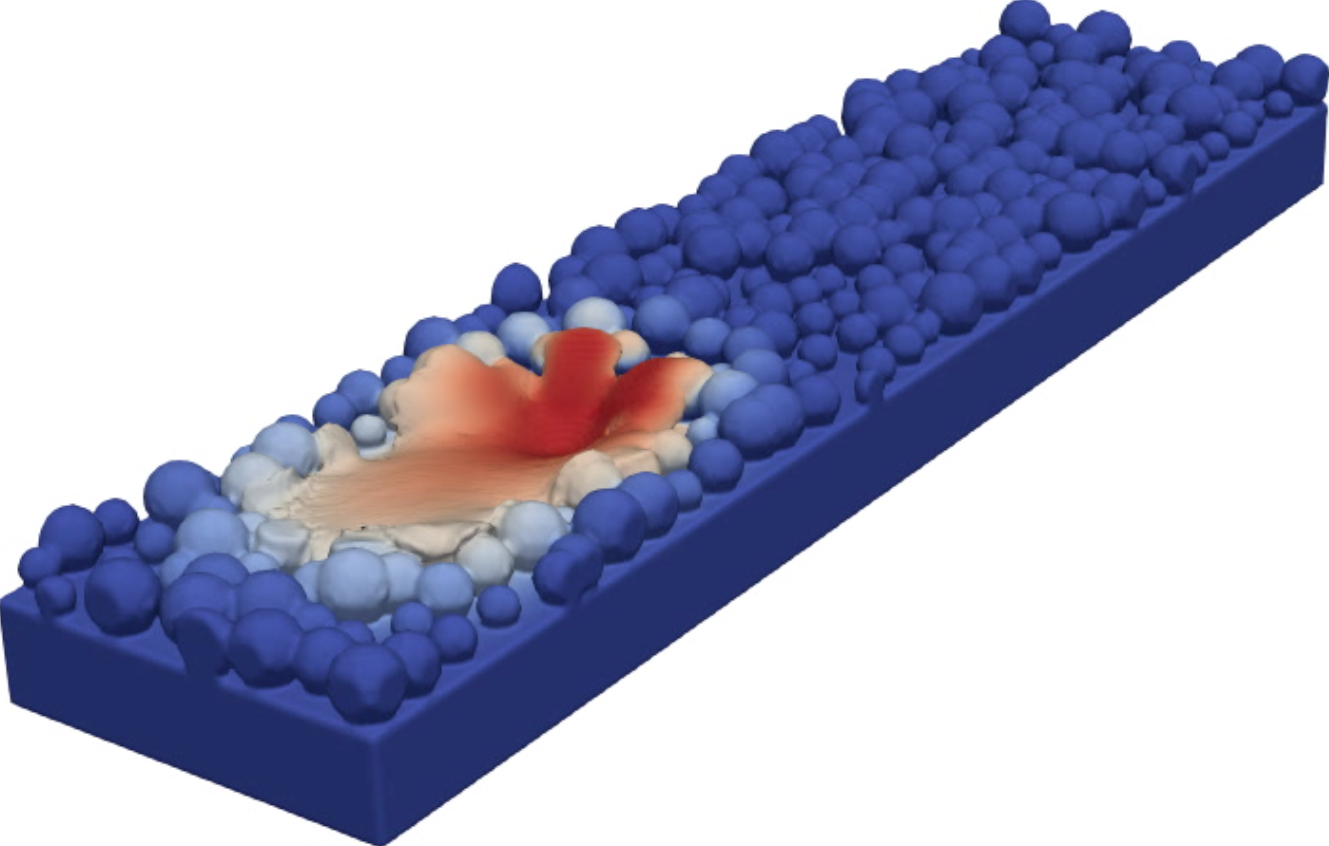}
\label{fig:example6_2a}
}
\subfigure [time: approx.  $0.48ms$ \hspace{0.22\textwidth}]
{
\includegraphics[width=0.48\textwidth]{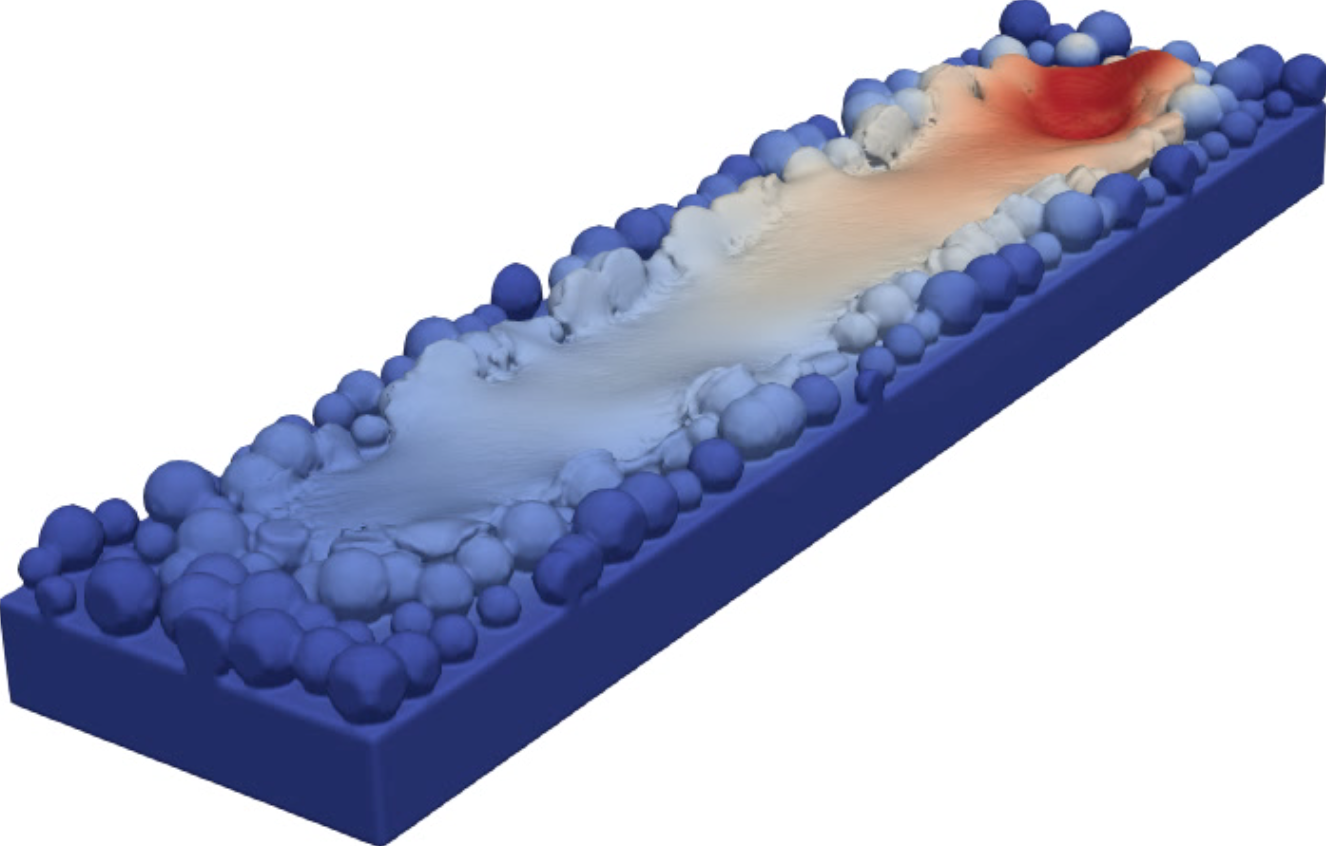}
\label{fig:example6_2d}
}
\caption{3D line melting problem with explicitly resolved powder particle geometry: resulting melt pool shape and final topology of solidified surface (post-processed) as well as temperature field in the range from $300K$ (blue) to $3700K$ (red)~\cite{Meier2020}.}
\label{fig:example6_2}
\end{figure}

Eventually, Figure~\ref{fig:example6_2} displays two configurations of a 3D line scanning example with spatially resolved powder particle geometry. Accordingly, surface tension forces dominate volumetric forces such as gravity at the considered length scales and smoothen out the original particle contours almost immediately after melting. The peak temperatures at the melt pool center exceed the boiling temperature, and the resulting recoil pressure forces foster the creation of a depression at the melt pool center.

\subsection{Alternative approach: High-fidelity melt pool model with resolved evaporation dynamics}
\label{sec:melt_pool_evaporation}

\noindent
{In order to describe mechanisms of defect creation, e.g., creation of gas inclusions as discussed above, more accurately and to generally consider evaporation-induced gas/vapor flows and thereby induced material re-distribution dynamics (e.g., powder particle entrainment by gas flow), a model is required with spatially resolved gas and vapor phase as well as liquid-vapor phase transition. Our ongoing research work focuses on the development of such an alternative, high-fidelity melt pool model, which differs from the SPH model described above by the following main aspects: (i) evaporation, i.e., the phase transition from liquid to vapor/gas phase is explicitly resolved, and the associated (recoil) pressure jump and heat transfer across the liquid-gas interface follow consistently from the balances of mass, momentum and energy instead of employing the phenomenological models~\eqref{eq:SPH_evap_recoil} and~\eqref{eq:SPH_evap_heatloss}; (ii) an Eulerian discretization scheme (viz. the finite element method, FEM) is applied, which requires a tracking scheme (viz. the level set method) for the position of the (diffuse) liquid-gas interface; (iii) a truly incompressible instead of weakly incompressible flow is considered in the liquid and gas phase. For this approach, model equations~\eqref{eq:SPH_basic} have to be supplemented by a transport equation for the level set method that additional allows for the liquid-vapor phase transition. Moreover, the continuity equation~\eqref{eq:SPH_conti} takes on the form of a constraint equation in the truly incompressible case. The pressure can be identified as associated Lagrange multiplier, and the densities $\rho_l$ and $\rho_g$ of the liquid and gas phase are constant and a priory known material parameters, i.e., the equation of state~\eqref{eq:SPH_eos} is not needed anymore in this case. In sum, the problem is described by a set of six equations, namely the incompressibility constraint equation, the Navier-Stokes momentum equation (3 components), the energy equation, and the level set transport equation, for the six primary unknowns given by velocity $\vectorbold{u}$ (three components), pressure~$p$, temperature~$T$ and a level set function $\phi$. {The code implementation of the high-fidelity melt pool model\footnote{{MeltPoolDG; see description on \url{https://github.com/meltpooldg/meltpooldg}}} strongly relies on the software package adaflo~\cite{kronbichler2018fast} and the deal.II finite element library~\cite{arndt2020deal,arndt2021deal}.}

\begin{figure}[h!!!]
	\includegraphics[width=\textwidth]{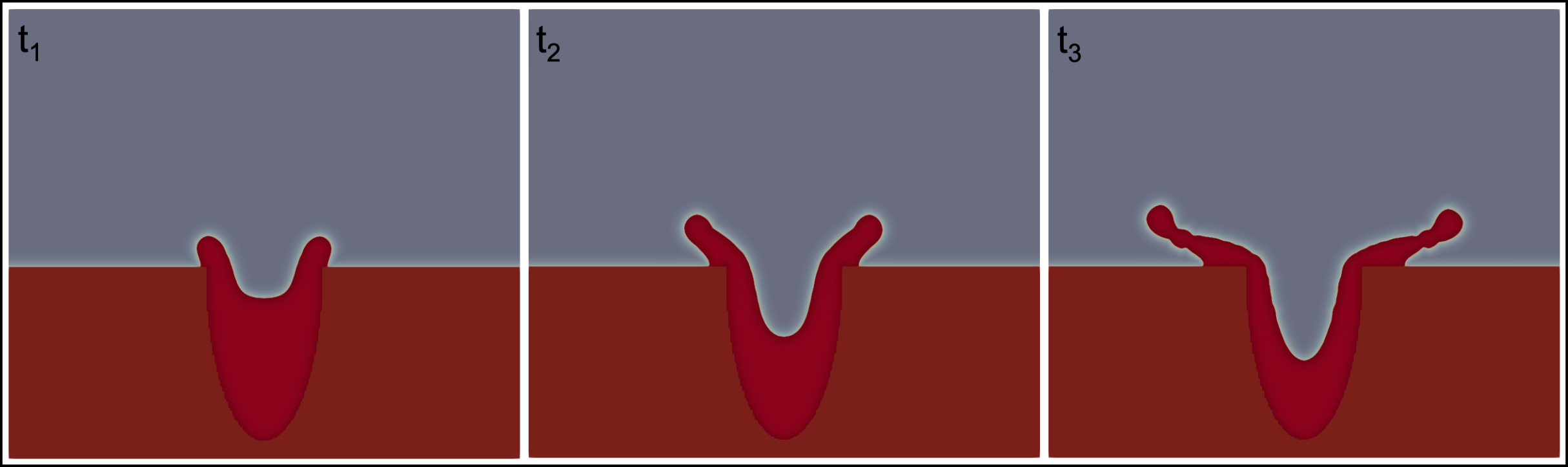}
	\caption{2D simulation of laser melting with \textit{phenomenological recoil pressure model}: The results show a contour plot of the resulting level set, which represents the melt pool shape, at three different time steps.}
	\label{fig:meltpool_recoil}
\end{figure}
\begin{figure}[h!!!]
	\includegraphics[width=\textwidth]{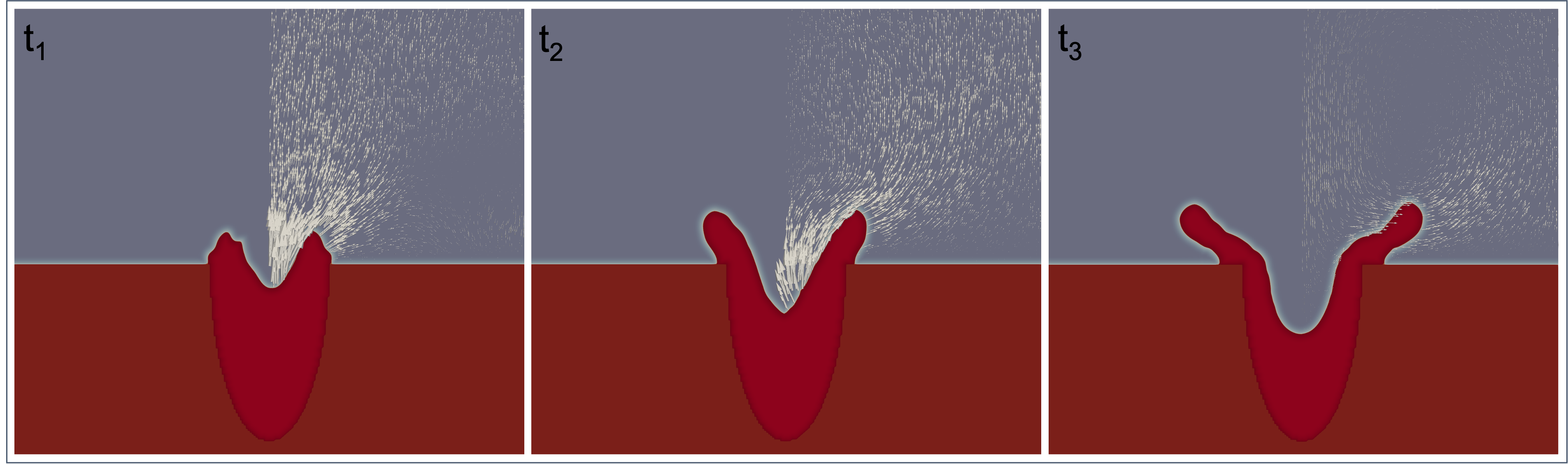}
	\caption{2D simulation of laser melting with \textit{consistent multi-phase evaporation model}: The results show a contour plot of the level set and velocity fields in the gaseous domain, the latter indicated by a vector plot, at three different time steps.}
	\label{fig:meltpool_evapor}
\end{figure}
Figures~\ref{fig:meltpool_recoil} and~\ref{fig:meltpool_evapor} represent preliminary simulation results of this high-fidelity melt pool model. Specifically, the variant with consistent multi-phase modeling and explicitly resolved liquid-vapor phase transition (see Figure~\ref{fig:meltpool_evapor}) is compared to a variant of this FEM model, where the phenomenological recoil pressure model~\eqref{eq:SPH_evap_recoil} is employed instead. For simplicity, and to isolate the effect of interest, the temperature field (relying on an approximate analytical solution) and the resulting melt pool shape are prescribed a priori for this example instead of solving the energy equation. Moreover, at the current state of the ongoing model development, the entire gaseous phase is assumed to consist of metal vapor, i.e., no local phase fractions of vapor and inert gas with correspondingly averaged material properties are considered. Please note also, that the parameters of the two variants have been chosen to lie in a comparable range, but due to the current state of model development a direct quantitative comparison of the results is not possible. Given these restrictions, the melt pool shapes in Figures~\ref{fig:meltpool_recoil} and~\ref{fig:meltpool_evapor} are in a reasonable qualitative agreement. For both variants, the evaporation-induced recoil pressure results in a deep depression at the center of the melt pool and melt spatter ejected from the pool at its lateral boundaries. Per underlying model assumptions, only the high-fidelity variant according to Figure~\ref{fig:meltpool_evapor} can predict the vapor velocity field resulting from the evaporation dynamics (visualized by white arrows).}

\subsection{Experimental validation}
\label{melt_exp}

{Again, drawing conclusions from the computational melt pool model requires experiment-based calibration, i.e. inverse identification, of the model parameters and validation of results. {These critical steps will be performed as part of our future research work based on the expertise of the MIT team and their infrastructure in terms of process control and metrology instrumentation following two principle routes -- in-situ optical process monitoring and ex-situ, post-process analyses.}}

{Optical monitoring techniques of PBFAM have been successfully employed to detect defect creation and correlate process signatures to final part properties. For example, defects, spatter, melt track dimensions and even melt pool instabilities can been detected \emph{in-situ} by identifying temperature developments in the laser scan path~\cite{Clijsters2014, Thombansen2014Tracking, Bisht2018, Coek2019,Thombansen2016, Rombouts2006,Fox2017,Yadroitsev2014, Doubenskaia2015, krauss2012thermography, krauss2014layerwise,Lane2016}. The team at MIT has developed a mid-wave infrared (MWIR) imaging technique, that overcomes limitations of previous techniques such as distortions and chromatic aberrations due to the usage of \textit{f}-$\theta$ lenses or perspective distortion and limited depth of field due to the typical off-set viewing position~\cite{Thombansen2014Lens,Craeghs2011,Craeghs2012}. The technique provides a high-resolution temperature field by capturing thermal emissions from the build surface. It applies a unique aperture division multiplexing (ADM) method, where the input aperture of a single large-diameter lens is sub-divided to bring the beam paths of the laser and the optical monitoring signal to focus on the build platform~\cite{penny2020additive}.}

{Calibration of model parameters can be conducted with e.g., simple point/line heating and melting experiments with varying laser powers. This can provide a good basis for model parameters such as laser absorptivity of the melt surface or temperature-dependence of the surface-tension, since the thereby induced Marangoni flow will considerably influence the temperature distribution in the melt pool. Also for the high laser power regime, where evaporation has considerable influence on the temperature profile and the generation of spatter, which is also identifiable via MWIR, point melting experiments are considered suitable to identify model parameters associated with the thermo- and fluid-dynamical phenomena of evaporation. Of course, process temperature signatures are also suitable for validation of actual powder melting simulations.}

{As second means for validation, ex-situ data will be considered. For post-process characterization, laser confocal microscopy of the sample surface, cross-sectional optical microscopy and scanning electron microscopy (SEM), as well volumetric imaging via micro-computed tomography (CT) can be employed. By this means, melt track morphology (e.g., shape, height, width), defects such as voids (size, shape, position, frequency), and surface topology can be used as metrics for model validation.}

\section{Macroscale Thermo-Solid-Mechanical Modeling} \label{sec:macro_modeling}

\noindent
{The TUM authors recently proposed a thermo-mechanical finite element model aiming at the prediction of residual stresses and thermal distortion in the partscale simulation of metal PBFAM processes~\cite{Proell2021,Proell2020}. The most important model equations as well as exemplary results are recapitulated in the following.}

{An alternative thermo-mechanical finite element model for PBFAM partscale simulations has been proposed by N. Hodge and the team at LLNL~\cite{Hodge2016,Hodge2014}. In particular, the PBFAM example in the microstructure studies of Section~\ref{sec:micro_results} has been simulated using this model and its code implementation in the in-house parallel finite element code \emph{Diablo}~\cite{diablo_manual_2018} at LLNL. Recent extensions of the underlying numerical schemes focus on the critical aspects of computational efficiency and accuracy by adequately addressing the spatial and temporal multiscale nature of these processes~\cite{hodge_towards_2020,ganeriwala_towards_2020}. For further details on this alternative partscale modeling approach the interested reader is referred to the aforementioned references.}

\subsection{Model equations}
\label{sec:macro_balance_equations}

\noindent
The thermo-mechanical problem consists of the dynamic heat equation coupled with the static balance of linear momentum:
\begin{align}
    \label{eq:heat_equation}
    c_p(T)\, \dot{T} +\nabla \cdot \vectorbold{q} &= \hat{r}, \\
    \label{eq:balance_linear_momentum}
    \nabla \cdot \boldsymbol{\sigma} &= \vectorbold{0},
\end{align}
with the primary variables temperature $T$ and displacement $\vectorbold{d}$. The magnitudes of strains and rotations arising from typical PBFAM processing conditions can be assumed as small, hence commonly the theory of linear continuum mechanics in combination with the engineering strain tensor
\begin{align}
    \label{eq:kinematics_epsilon}
    \boldsymbol{\varepsilon} = \frac{1}{2}\left(\nabla \boldsymbol{d} + (\nabla \boldsymbol{d})^{T}\right)
\end{align}
as metric for deformation are employed. The structural material law relating those with the Cauchy stresses $\boldsymbol{\sigma} = \boldsymbol{\sigma}(\boldsymbol{\varepsilon}(\vectorbold{d}), T)$ will be detailed Section~\ref{sec:macro_mechanical_law}. The heat flux $\vectorbold{q}$ is specified by Fourier's law of heat conduction,
\begin{align}
    \vectorbold{q} = -k(T) \nabla T.
\end{align}
The material parameters appearing in the thermal problem, namely volumetric heat capacity $c_p$ and heat conductivity $k$, in general depend on temperature and phase as discussed in the following Section~\ref{sec:macro_temp_phase_params}. Finally, $\hat{r}$ represents a volumetric laser beam heat source, which is modeled according to~\cite{Gusarov2009}.

\subsubsection{Temperature- and phase-dependent parameters}
\label{sec:macro_temp_phase_params}

\noindent
To model the three relevant phases powder, melt and solid, in a first step the liquid phase fraction $g$ is defined according to
\begin{align}
    \label{eq:liquid_fraction}
    g(T) = \begin{cases}
        0, & T < T_s\\
        \frac{T-T_s}{T_l-T_s}, &T_s \leq T \leq T_l\\
        1, &T > T_l,
    \end{cases}
\end{align}
where $T_l$ and $T_s$ represent the liquidus and solidus temperature. In a next step, the irreversibility of the powder-to-melt transition during melting is captured via the consolidated fraction
\begin{align}
    \label{eq:consolidated_fraction}
    r_c(t) = \begin{cases}
        1, & \text{if } r_c(t=0)=1 \text{ (i.e., initially consolidated)}\\
        \underset{\tilde{t} \leq t}{\max}\, g(T(\tilde{t})), & \text{if } r_c(t=0)=0 \text{ (i.e., initially powder)}.\\
    \end{cases}
\end{align}
In a last step, the final phase fractions of the powder ($p$), melt ($m$) and solid ($s$) material are computed following the relations:
\begin{align}
    r_p &= 1 - r_c,\\
    r_m &= g,\\
    r_s &= r_c -g.
\end{align}
Eventually, these phase fractions can be used to interpolate an arbitrary material parameter $f$ according to the following scheme:
\begin{align}
    \label{eq:material_parameter_interp}
    f_\text{interp} = r_p(T) f_p(T) + r_m(T) f_m(T) + r_s(T) f_s(T),
\end{align}
where $f_\text{interp}$ is the interpolated parameter value and $f_p$, $f_s$ and $f_m$ are the single phase parameters. This approach is applied to the thermal conductivity $k$ and heat capacity $c$. The phase- and temperature-dependent formulation of mechanical material properties is presented in the next section.

\subsubsection{Mechanical constitutive law}
\label{sec:macro_mechanical_law}

\noindent
An iso-strain (Voigt-type) homogenization, assuming equal strains in all three phases, is applied in the following. Accordingly, the stress of the mixture is given by a weighted sum of the individual contributions, a procedure that is in fact similar to the interpolation scheme~\eqref{eq:material_parameter_interp}:
\begin{align}
    \label{eq:stress_total}
\boldsymbol{\sigma} = \sum_i {r}_i\,\boldsymbol{\sigma}_{i} \quad \text{with} \quad i \in \{p,m,s\}.
\end{align}
In general, the strain \eqref{eq:kinematics_epsilon} in a single phase $i \in \{p,m,s\}$ is calculated from an additive decomposition of the strain tensor
\begin{align}
    \label{eq:additive_split}
    \boldsymbol{\varepsilon} = \boldsymbol{\varepsilon}_i = \boldsymbol{\varepsilon}_{E,i} + \boldsymbol{\varepsilon}_{p,i} + \boldsymbol{\varepsilon}_{T,i} + \boldsymbol{\varepsilon}_{\text{ref},i},
\end{align}
although not all terms will be utilized for each phase. The total or kinematic strain on the left-hand side of~\eqref{eq:additive_split} is purely displacement-dependent and for all phases given by the kinematic relation~\eqref{eq:kinematics_epsilon}. The first term on the right-hand side of~\eqref{eq:additive_split} is the elastic strain $\boldsymbol{\varepsilon}_{E,i}$, which leads to a stress $\boldsymbol{\sigma}_i$ in each phase 
\begin{align}
\boldsymbol{\sigma}_i = \ctensor_i : \boldsymbol{\varepsilon}_{E,i}, 
\end{align}
when considering a linear hyper-elastic material with fourth-order elastic constitutive tensor $\ctensor_i$, e.g., according to Hooke's law with Young's modulus $E_i$ and Poisson's ratio $\nu$ (assumed to be equal in all phases) as independent constitutive parameters. The plastic strains $\boldsymbol{\varepsilon}_{p,i}$ are only relevant for the solid phase, and can be calculated using standard approaches, e.g., an incremental problem formulation in combination with a return mapping algorithm. For simplicity, this contribution will not be considered in the examples discussed below. The strains $\boldsymbol{\varepsilon}_{T,i}$ due to thermal expansion are assumed equal in all phases and read
\begin{align}
    \boldsymbol{\varepsilon}_{T,i} = \boldsymbol{\varepsilon}_{T}  = \boldsymbol{I}\int_{T_\text{ref}}^T\alpha_{T}\,\dd T =   \alpha_{T}(T-T_\text{ref}) \boldsymbol{I},
\end{align}
where $\alpha_{T}$ is the coefficient of thermal expansion. Critically, an inelastic reference strain $\boldsymbol{\varepsilon}_{\text{ref},i}$, which is only relevant for the solid phase, is proposed in rate form according to:
\begin{align}
    \label{eq:epsilon_ref_total_rs}
   \boldsymbol{\varepsilon}_{\text{ref}}=\frac{1}{r_s} \hat{\boldsymbol{\varepsilon}}_{\text{ref}} \quad \text{with} \quad \dot{\hat{\boldsymbol{\varepsilon}}}_{\text{ref}} = 
\begin{cases}
        (\boldsymbol{\varepsilon} - \boldsymbol{\varepsilon}_{p} - \boldsymbol{\varepsilon}_{T}) \cdot \dot{r}_s, \quad &\text{if } \dot{r}_s > 0\\
        {\hat{\boldsymbol{\varepsilon}}_{\text{ref}}\, \frac{\dot{r}_s}{r_s},}\quad &{\text{otherwise},}
  \end{cases}
  \quad \quad \text{and} \quad \hat{\boldsymbol{\varepsilon}}_{\text{ref}} (t=0) = \vectorbold{0}. 
\end{align}
Note, that the reference strains only change when the solid phase fraction increases according to $\dot{r}_s >0$ (first case in~\eqref{eq:epsilon_ref_total_rs}), i.e., for temperatures $T \in [T_s; T_l]$ in the phase change interval and negative temperature rates $\dot{T}<0$. {Note also that the second case in~\eqref{eq:epsilon_ref_total_rs} formally ensures that $\dot{\boldsymbol{\varepsilon}}_{\text{ref}}=0$ for $\dot{r}_s \leq 0$. In practice, however, ${\boldsymbol{\varepsilon}}_{\text{ref}}$ is simply kept constant in time intervals where no solid phase fraction is created (i.e., $\dot{r}_s \leq 0$) instead of integrating the evolution equation for $\dot{\hat{\boldsymbol{\varepsilon}}}_{\text{ref}}$.} An elastic constitutive law with low stiffness values (i.e., $E_p, E_m, \ll E_s$) as applied to powder and melt leads to small stresses yet considerable total strains in these phases. In this context, the reference strains according to~\eqref{eq:epsilon_ref_total_rs} ensure that these strains do not translate into stresses during solidification. For the special case that kinematic $\boldsymbol{\varepsilon}$, plastic $\boldsymbol{\varepsilon}_{p}$ and thermal strains $\boldsymbol{\varepsilon}_{T}$ are constant during solidification, which approximately holds if the phase change interval $T_l-T_s$ is sufficiently small, it can easily be verified from~\eqref{eq:additive_split} and~\eqref{eq:epsilon_ref_total_rs} that the elastic strain, and thus the resulting stresses, in the evolving solid phase vanish.\\

\begin{minipage}{16.5 cm}
\textbf{Remark:} One of the main assumptions underlying the present and most existing thermo-mechanical \nohyphens{PBFAM} models is that mechanical stresses in the (open-surface) powder and melt phase domains are negligible. This behavior is approximated by applying a simple elastic constitutive law to these phases, with stiffness parameters that are considerably lower as compared to the solid phase, i.e., $E_p, E_m \ll E_s$. In practice, this approximation turns out to result in moderate, i.e., limited, strains, since the deformation of these powder and melt domains is mostly kinematically controlled by the motion of the significantly stiffer solid phase domains, thus yielding only small stress contributions as desired. Moreover, as compared to approaches exactly satisfying the zero-stress assumption in powder and melt, no additional means (e.g., extended finite element method, immersed boundary method, etc.) are required for representation of discontinuities (e.g., jumps in stresses) inside elements or for mesh movement in the geometrically nonlinear case. Note, the assumption that thermal strains exist also in the powder and melt phase, and are equal to thermal strains in the solid phase, has been made for simplicity here. This assumption is neither necessary nor has it a significant influence on the resulting residual stresses due to the low stiffness of these phases and the definition of reference strains~\eqref{eq:epsilon_ref_total_rs}, which ensure that newly created solid material is stress-free.
\end{minipage}

\subsection{Exemplary simulation results} \label{sec:macro_results}

\noindent
In~\cite{Proell2021}, the accuracy of the proposed thermo-mechanical model and the influence of the individual model components has been critically verified by means of elementary test cases, partly with analytical solutions. 

\begin{figure}[t!!!!]
    \centering
    \includegraphics[width=\linewidth]{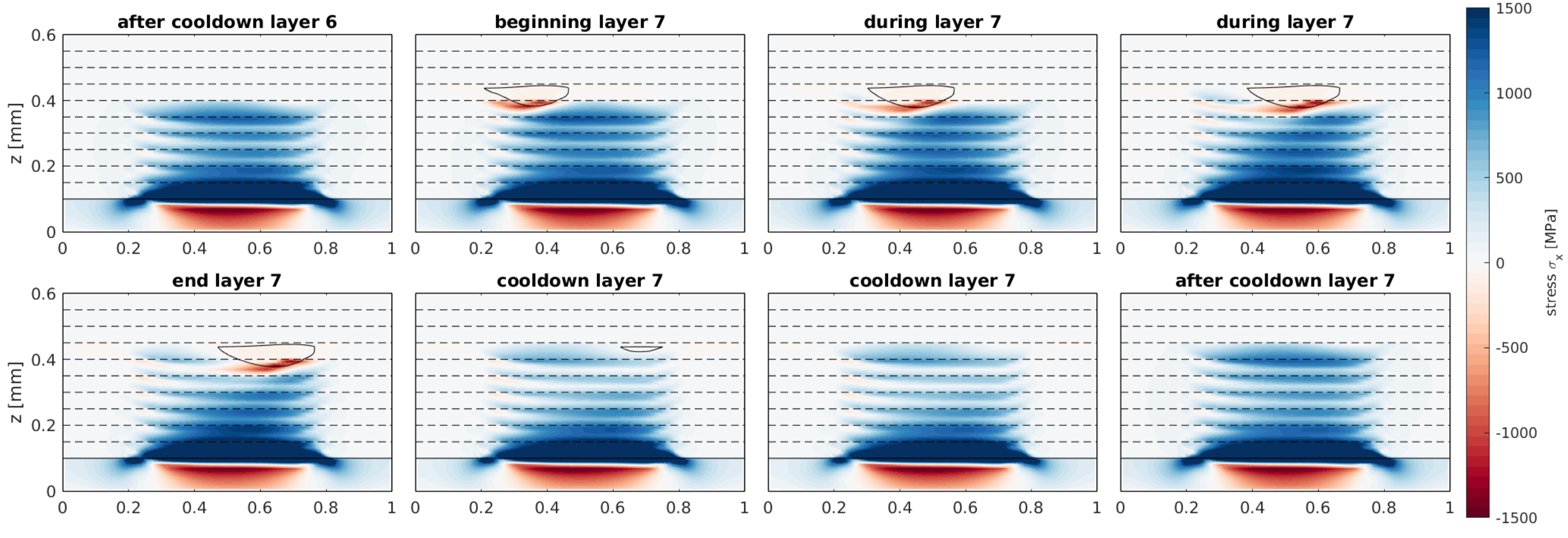}
    \caption{Detailed view of the evolution of the normal stress $\sigma_{x}$ in horizontal direction during scanning of layer 7. Melt pool indicated with black contour line. Time progresses from left to right and top to bottom~\cite{Proell2021}.}
    \label{fig:10layer_stressx_detail_layer7}
\end{figure}

In the following, two examples with direct relevance for macroscale modeling of PBFAM processes will be briefly recapitulated. The first example consists of a solid base plate (confined by horizontal solid line) and ten melt tracks (separated by dashed horizontal lines) successively deposited on top of each other. Here, the length of the (horizontally centered) melt tracks has been chosen as half of the length of the base plate, and the laser is scanning from left to right in each track. The resulting normal stresses in x-direction, i.e., in horizontal direction, for different snapshots in time are illustrated in Figure~\ref{fig:10layer_stressx_detail_layer7}. As characteristic feature of the stress distribution, a horizontal band pattern can be identified, with stresses alternating between positive (tensile) stress values in the upper part and stress values close to zero in the lower part of a each melt track. In the following, the creation of this track-wise band pattern will be explored exemplarily during processing of track $7$. 
\begin{figure}[t!!!]
\centering
%
\subfigure [Equivalent von Mises stress after final cooldown.]
{
\includegraphics[width=0.48\textwidth]{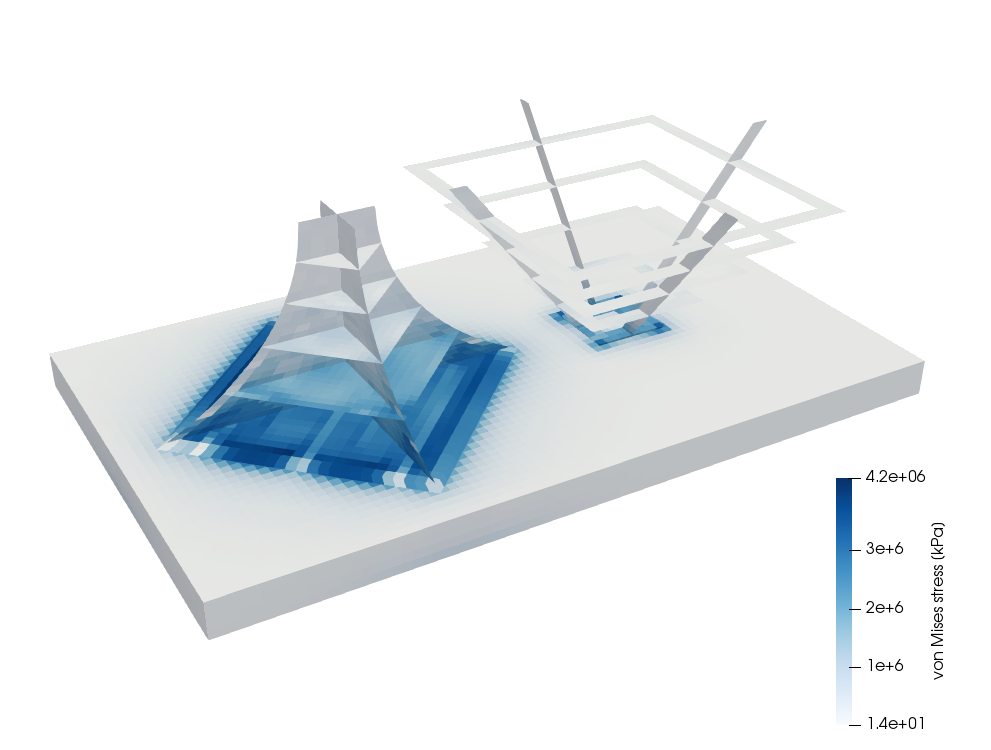}
\label{fig:doublepyramit_1a}
}
\subfigure [Displacement magnitude after final cooldown displayed on warped geometry. The distortion is scaled up by a factor of 5.]
{
\includegraphics[width=0.48\textwidth]{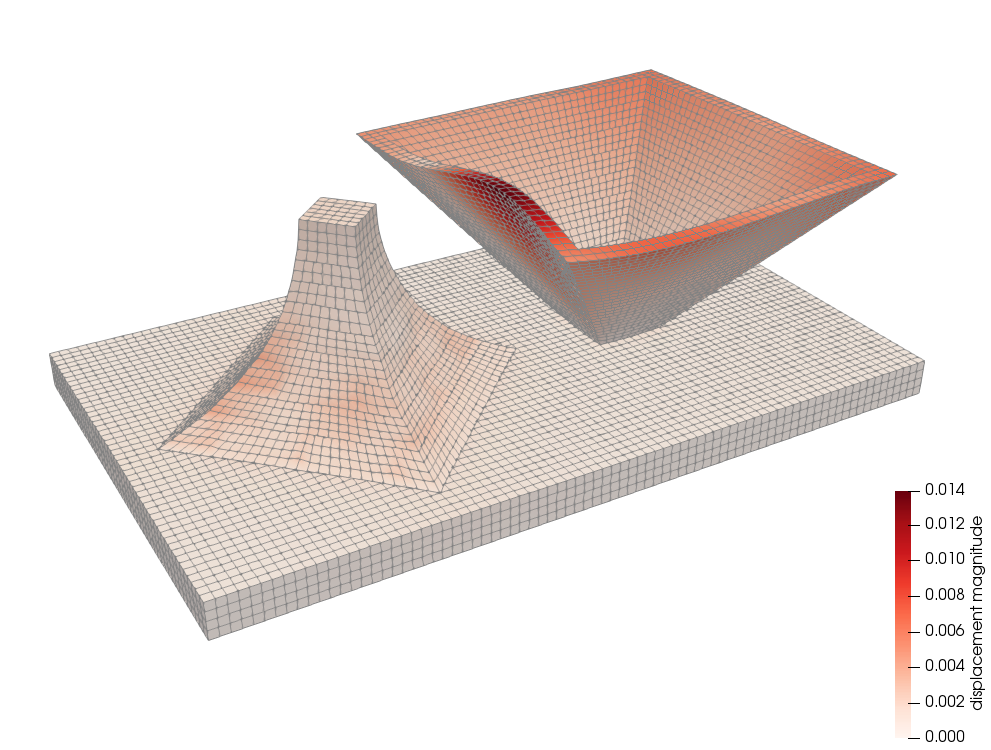}
\label{fig:doublepyramit_1b}
}
\caption{Processing of a solid and hollow pyramid. Discretization with non-matching meshes via mortar mesh-tying~\cite{Proell2021}.}
\label{fig:doublepyramit_1}
\end{figure}
The top left snapshot of Figure~\ref{fig:10layer_stressx_detail_layer7} represents a configuration, where track $6$ has been finished and cooled down already. The subsequent snapshots (from left to right) in the first and second row of Figure~\ref{fig:10layer_stressx_detail_layer7} show the deposition of track $7$. To visualize the melt pool size, the temperature isoline corresponding to the solidus temperature $T_s$ is depicted in these snapshots (solid black line). While no stresses occur in the powder material of layer $7$ in front of the laser, the thermally induced volume expansion during heating leads to negative (compressive) stresses in the solid material of the previously processes track $6$, mostly pronounced in the direct vicinity of the $T_s$-isoline. As desired, these stresses rapidly drop to zero in the narrow temperature interval $T \in [T_s;T_l]$ such that no visible stresses remain in the melt pool domain. This strong gradient between vanishing stresses in the melt pool and high compressive stresses in the solid material beneath remains after solidification and is superimposed by additional tensile stress contributions due to the thermally induced volume shrinkage during cooling. After cool down (see snapshot at bottom right of Figure~\ref{fig:10layer_stressx_detail_layer7}), this process results in high tensile stresses in the upper, re-molten part of track $6$, and stresses close to zero in its lower part. The same characteristics are observed in the previously processed tracks below. Even though the base plate has the same stiffness and similar support conditions (i.e., spatially fixed only at its bottom surface) as the solidified tracks above, this characteristic temperature gradient is much more pronounced for the first track, i.e., the highest overall tensile stresses occur in the first track, accompanied by compressive stresses of comparable magnitude in the base plate below. This observation can be explained by the fact that the base plate is initially stress-free, while solidified melt tracks are subject to tensile stresses after cooling down (e.g., snapshot at bottom right of Figure~\ref{fig:10layer_stressx_detail_layer7}), which partly compensate the compressive stresses arising from thermal expansion when processing the subsequent track above.

As important feature of the proposed macroscale modeling framework, successive layers of material are connected by powerful mortar mesh-tying schemes~\cite{popp2012dual, puso2003mesh}. These schemes allow for non-matching meshes between these layers while preserving optimal convergence rates in the $L^2$-norm, which is typically not guaranteed by alternative, e.g., collocation-type, mesh-tying approaches. In particular, the combination of complex geometries, e.g., tapered parts with strongly changing cross-section dimensions, and the need for layer-wise mesh definition makes the discretization of PBFAM parts challenging when using standard discretization schemes. In Figure~\ref{fig:doublepyramit_1}, a geometry of this type is illustrated, consisting of a solid pyramid with curved edges and an upside-down hollow pyramid. Note that the meshes used for discretization of the pyramids are non-matching between successive layers, and also non-matching between the pyramids and the base plate. The flexibility arising from this non-matching discretization approach allows to have very regular meshes with undistorted and almost equally-sized finite elements, which does not only decrease the effort and time for mesh generation but also the number of finite elements, i.e., degrees of freedom, to achieve a certain approximation quality, i.e., to lie below an acceptable discretization error level.

Again, the visualization of von Mises stresses in Figure~\ref{fig:doublepyramit_1a} reveals that highest residual stresses occur in the first layer above the base plate. The displacement magnitudes associated with the coupled thermo-mechanical problem are illustrated in Figure~\ref{fig:doublepyramit_1b}. {Accordingly, the thin walls of the hollow pyramid exhibit a strong thermal distortion during cool down}.

\vspace{-0.03 \textwidth}

{\subsection{Experimental validation}}
\label{macro_exp}

{Our future research work will focus on experimental validation of the macroscale model. Quantification of residual stress and distortion, respectively, require measurements of the lattice distortion of solidified material and the three-dimensional component shape. Both are intrinsically challenging, and therefore much research has relied on fabrication and metrology of geometric test artifacts. As such, early work by Kruth and colleagues studied the influence of scan pattern (e.g., unidirectional vs. raster scan) on residual stress development in small arch-shaped geometries~\cite{mercelis2006residual}. The coupled influence of support structure placement and geometry on deformation and residual stress has also been studied. In situ measurements of stress buildup in LPBF were performed using a build plate instrumented with strain sensors~\cite{dunbar2016development}, or, alternatively, by embedding strain-sensing fiber optics in SLM components~\cite{havermann2015measuring}. High energy X-ray and neutron diffraction scattering enable local, volumetric probing of residual stress, and can be employed ex situ (e.g., after printing) as well as during heat treatment.  The sample geometry must ensure beam penetration; neutron diffraction is attractive due to the higher penetration depth through metallic materials~\cite{goel2020residual}, where X-ray techniques are preferred for thin film geometries. Last, three-dimensional imaging of surface and shape, such as by digital image correlation (e.g.,~\cite{bartlett2018revealing}) and computed tomography can be compared to simulations of component deformation.}

{
Foundational work by N. Hodge and the team at LLNLHodge~\cite{hodge2016experimental} demonstrates tight agreement between a continuum model and residual stress measurements by digital image correlation and neutron diffraction. Techniques for stress mitigation and active control are especially necessary when fabricating components where geometric accuracy of as-printed features or predictable stress-strain behavior is paramount, both in LPBF and directed energy deposition (DED) methods.}\\

\section{Microstructure Modeling} \label{sec:micro_modeling}

\noindent
{The authors recently proposed a microstructure model for phase fraction prediction in PBFAM of \nohyphens{Ti-6Al-4V}~\cite{Meier2020_2}. The most important model equations and exemplary results are recapitulated in the following.}

\subsection{Model equations}
\label{sec:micro_equations}

\noindent
Instead of spatially resolving individual crystal or grain geometries, the model describes solid state transformations in Ti-6Al-4V via spatially homogenized phase fractions of the most relevant metallurgical specifies, namely the $\beta$-phase, the stable $\as-$phase as well as the metastable martensitic $\am$-phase. As compared to spatially resolved approaches, this procedure offers the general suitability for partscale PBFAM simulations. The temperature field, which is required as input for the microstructure model, can e.g., be provided by macroscale thermal or thermo-mechanical PBFAM models according to Section~\ref{sec:macro_modeling}. The composition of the solid phases $\beta$, $\as$ and $\am$ are described by phase fractions $\mathrm{X}_i \in [0;1]$ fulfilling
\begin{subequations}
\label{eqn: continuity_constraints}
\begin{align}
    \Xalpha+\Xb &=1,\label{eqn: continuity_constraints2}\\
    \Xa + \Xm &= \Xalpha,\label{eqn: continuity_constraints3}
\end{align}
\end{subequations}
for fully solidified material. Here, the stable and martensitic $\alpha$ phases, $\as$ respectively $\am$, are combined to the total alpha phase fraction $\Xalpha$.
For sufficiently slow cooling, only the stable phases $\as$ and $\beta$ will arise. The temperature-dependent equilibrium phase fraction~$\Xaeq(T)$, towards which the $\as$-phase tends in the extreme case of very slow cooling rates~$|\dT| \ll |\dTm|$, is described as Koistinen-Marburger law~\cite{koistinen} according to: 
\begin{align}
\label{eqn: mean_approx}
\begin{split}
 \Xaeq(T)&=
\begin{cases}
0.9&\text{for }T<\Tbs,\\
1-\exp\left[-\krateeq \cdot \left(\Tbe-T\right)\right]&\text{for }\Tbs\le T\le \Tbe,\\
\end{cases}
\end{split}
\end{align}
Thus, for temperatures above the alpha-transus end temperature $\Tbe=935K$ pure $\beta$ material, and for temperatures below the alpha-transus end temperature $\Tbs=848K$ a phase composition with $10\%$ $\beta$- and $90\%$ stable $\alpha_s$-phase are expected under thermodynamic equilibrium conditions. For the second extreme case of very fast cooling rates $|\dT| \geq |\dTm|$, at which the diffusion-driven formation of $\Xa$ is suppressed, a \textit{metastable} Martensite pseudo equilibrium $\Xmeqo(T)$ is stated~\cite{Fan.2005,koistinen,Gil.1996}: 
\begin{align}
\label{eqn: am_equo}
\begin{split}
 \Xmeqo(T)&=
\begin{cases}
0.9&\text{for }T<T_\infty,\\
1-\exp\left[-\kratem(\Tms-T)\right]&\text{for }T_\infty \le T \le \Tms,\\
0&\text{for }T>\Tms.
\end{cases}
\end{split}
\end{align}
Here, $|\dTm|$ is denoted as critical cooling rate, above which pure Martensite formation is observed. Note that this critical rate is not prescribed in our model but emerges naturally from the system dynamics. Note also that, according to this model, Martensite formation can only be initiated for temperatures below the Martensite start temperature $\Tms$, and a maximal Martensite phase fraction of $90\%$ can be achieved at room temperature $T_\infty=293K$. Finally, in the most general case of cooling rates that are too fast to complete the diffusion-driven $\as$-formation before reaching the Martensite start temperature $\Tms$ but still below the critical rate $|\dTm|$, i.e., a certain amount of stable $\as$-phase has already been formed at $\Tms$, a Martensite phase fraction below $90\%$ is expected at room temperature. For this case, we propose to replace~\eqref{eqn: am_equo} by an effective pseudo equilibrium phase fraction $\Xmeq(T)$ accounting for the reduced amount of transformable $\beta$-phase at presence of a pre-existing phase fraction $\Xa$:
\begin{align}
\label{eqn: am_eq_general}
 \Xmeq(T)= \Xmeqo(T) \cdot \frac{(0.9-\Xa)}{0.9}.
\end{align}
Eventually, the formation and dissolution of the three species is described in rate form, i.e., evolution equations are proposed with the following contributions to the total rates~$\dXa$, $\dXm$ and $\dXb$:
\begin{subequations}
\label{eqn: definition_rates}
\begin{align}
    \label{eqn: definition_rates_alphas}
    \dXa & = \dXbtoa + \dXmtoa - \dXatob,\\
    \label{eqn: definition_rates_alpham}
    \dXm &= \dXbtoam - \dXmtoa - \dXmtob,\\
    \label{eqn: definition_rates_beta}
    \dXb &= \dXatob + \dXmtob - \dXbtoa - \dXbtoam.
\end{align}
\end{subequations}
Here, e.g., $\dXbtoa$ represents the formation rate of $\as$ out of $\beta$ while $\dXatob$ represents the dissolution rate of $\as$ to $\beta$. Since the $\beta$-phase fraction can be calculated from the continuity constraint~\eqref{eqn: continuity_constraints}, the first two rate equations in~\eqref{eqn: definition_rates} are sufficient to predict the evolution of $\Xa$ and $\Xm$. Based on the physical mechanism underlying the phase transformation, the formation and dissolution rates in~\eqref{eqn: definition_rates} are classified as instantaneous vs. diffusion-based transformations, which are initiated through deviations of the current martensitic and total $\alpha$-phase fraction $\Xm$ respectively  $\Xalpha=\Xa + \Xm$ from the corresponding equilibrium values according to~\eqref{eqn: am_eq_general} and~\eqref{eqn: mean_approx}. While  instantaneous transformations are modeled via constraint equations enforcing that phase fractions follow the associated equilibrium value, the diffusion-based creation of a phase $B$ out of a phase $A$ is modeled as a modified logistic differential equation~\cite{avramov2014generalized} of the following form:
\begin{equation}
    \label{eqn: alpha_formation_der_new}
    \dot{X}_{B \rightarrow A} = 
    \begin{cases}
        k_{A}(T) \cdot \left(X_A \right)^{\frac{c_{A}-1}{c_{A}}} \cdot \left(X_B - X_B^{eq} \right)^{\frac{c_{A}+1}{c_{A}}}& \text{for }\  X_B > X_B^{eq},\\
        0&\text{else}.
    \end{cases}
\end{equation}
Here, the factor $(X_B - X_B^{eq})$ represents the driving force of the diffusion process in terms of transformable $B$-phase and has a decelerating effect on the transformation during the ongoing diffusion process. The factor with $X_A$ leads to a transformation rate that increases with increasing amount of created $\as$-phase, i.e., it has an accelerating effect on the transformation rate during the ongoing diffusion process. The factor $k_{A}(T)$ represents the temperature-dependent diffusion rate of this thermally activated process, taking into account the temperature-dependent mobility of the diffusing species. Eventually, the choice of the exponent $c_{A}$ dependents on the specific physical mechanisms underlying the diffusion process. Specifically, the rates $\dXbtoam$ for Martensite formation out of $\beta$ and $\dXmtob$ for Martensite disolution into $\beta$ are modeled as instantaneous transformations, the remaining rates in~\eqref{eqn: definition_rates_alphas} and~\eqref{eqn: definition_rates_alpham} as diffusion-based processes according to~\eqref{eqn: alpha_formation_der_new}. Further details are given in~\cite{Meier2020_2}.


\subsection{Exemplary simulation results}
\label{sec:micro_results}

\noindent
In~\cite{Meier2020_2}, time temperature transformation (TTT) experiments are used for model calibration,
    \begin{figure}[b!!!]
        \centering
        \vspace{-2.0mm}
        \includegraphics[scale=0.35]{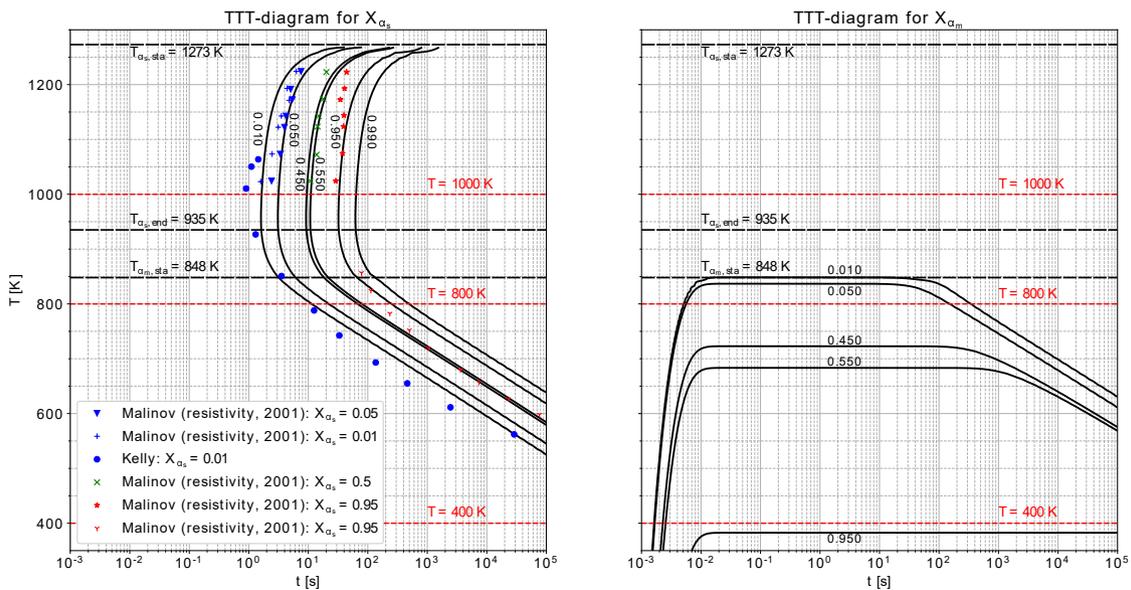}
        \vspace{-2.0mm}
         \caption{Simulation of TTT-diagram for the $\as$- and $\am$-phase, along with experimental data by Malinov~\cite{Malinov.2001.Resistivity} and Kelly~\cite{Kelly.article}. Left: Contour-lines for $\Xa$; Right: Contour-lines for $\Xm$. Contour lines are shown for the $1\%$, $5\%$, $45\%$, $55\%$, $95\%$ and $99\%$ normalized phase fractions. Three temperatures are marked in red and discussed in the analysis~\cite{Meier2020_2}.} 
        \label{fig: TTT_diagram}
    \end{figure}
    i.e., to inversely determine the unknown model parameters (e.g., in the diffusion equation~\eqref{eqn: alpha_formation_der_new}) such that the deviation between phase fraction contour-lines from experiments and corresponding simulations are minimized in a least-square sense. In TTT experiments, the material is first equilibriated at high temperatures such that only the high-temperature phase, here the $\beta$-phase, is present. Subsequently, the material is rapidly cooled down to a target temperature at which it is then held constant over time so that the isothermal phase transformation at this temperature can be recorded. Rapid cooling refers here to a cooling rate that is fast enough so that diffusion-based transformations during the cooling itself can be neglected and can subsequently be studied under isothermal conditions at the chosen target temperature. The procedure is repeated for successively reduced target temperatures. The emerging diagram of phase contour-lines over the $T\times\log(t)$ space is commonly referred to as TTT-diagram, which has been plotted for experimental measurements and model predictions of $\Xa$ and $\Xm$ in Figure~\ref{fig: TTT_diagram}. Note that the depicted phase fraction values are normalized to the corresponding equilibrium value. In addition to the TTT experiments used for model calibration, continuous-cooling transformation (CCT) experiments have been considered in~\cite{Meier2020_2} for model verification. In CCT experiments, the microstructural probe is again equilibriated at high temperatures where only the high-temperature $\beta$-phase is present. Afterwards, the probe is cooled down to room temperature $T_{\infty}=293.15\ K$ at different cooling rates $\dot{T}_{\text{CCT}}$ following precisely defined, time-continuous temperature profiles. Eventually, the evolving microstructure resulting from this cooling procedure is recorded on the $t\times T$-space. It was demonstrated that the model prediction of the critical cooling rate $|\dTm| \approx 410K/s$ above which pure Martensite formation was observed, is in very good agreement to experimentally measured values. In many existing models this critical cooling rate is \textit{enforced} as ad-hoc criterion for Martensite formation. In our model this characteristic cooling rate is not prescribed but is a direct \textit{consequence} of the energy and mobility competition between the microstructural species.
 \begin{figure}[htbp]
 \centering
    \includegraphics[width=1\linewidth]{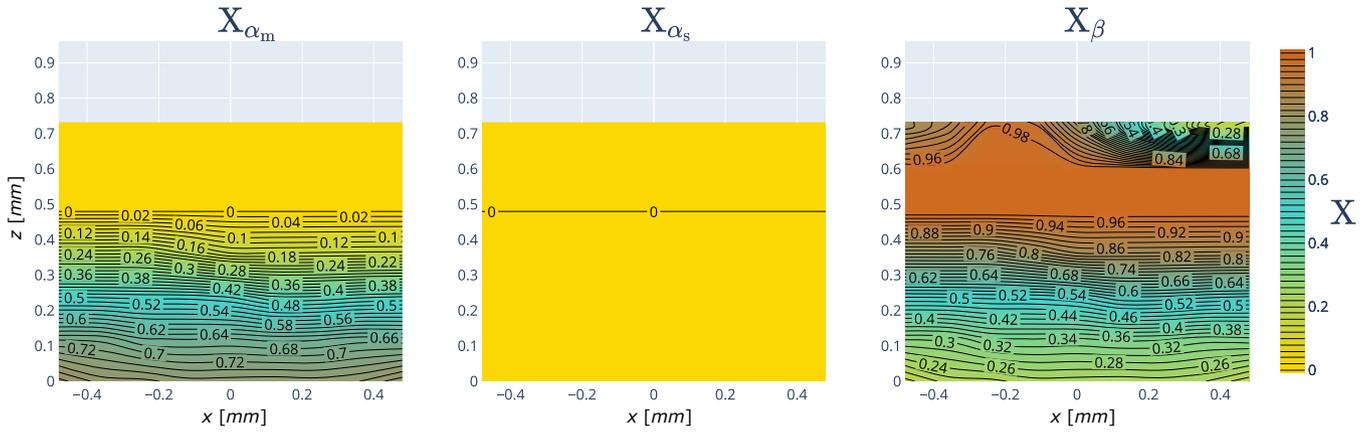}
    \caption{Simulated microstructure composition within the vertical center plance of a one millimeter-sided cube fabricated with SLM. The gray area above marks the layers that are not yet processed by the laser at the considered time. In the right figure, which depicts the $\beta$-phase fraction $\Xb$, the melt-pool is indirectly visible (the liquid phase fraction is not depicted) through the decreased $\beta$-phase fraction in the right corner of the up-most layers due to the laser that has just previously scanned this plane in x-direction from left to right. In a similar fashion, the decreased amount $\Xb$ in the upper left corner of the right Figure stems from the heat of the laser that is already melting the subsequent track at the considered time~\cite{Meier2020_2}.}
    \label{fig: plane_303K3s}
 \end{figure}
 
{As first PBFAM-specific example, Figure~\ref{fig: plane_303K3s} shows the simulated microstructure composition in the vertical center plane of a cube fabricated by selective laser melting. For this specific problem, the required thermal field has been provided by a macroscale PBFAM model developed and implemented at the LLNL~\cite{Hodge2016,Hodge2014}. To accurately predict the microstructure evolution in PBFAM, it is necessary to spatially resolve the correct laser path and layer/track dimensions instead of applying layer/track up-scaling or agglomeration approaches~\cite{Zaeh2010,Hodge2016,Zhang2018a}. Since the partscale simulation of PBFAM with resolved laser path is still an open research question, the following example will be limited to a cube of side length 1mm consisting of 34 layers.} According to Figure~\ref{fig: plane_303K3s}, no stable $\alpha_s$-phase is formed due to the fast cooling rates in PBFAM. Instead, the $\beta$-phase directly transforms into martensitic $\am$-phase once the temperature falls below the Martensite start temperature $\Tms$.
 \begin{figure}[t!!!]
  \centering
    \includegraphics[scale=0.48]{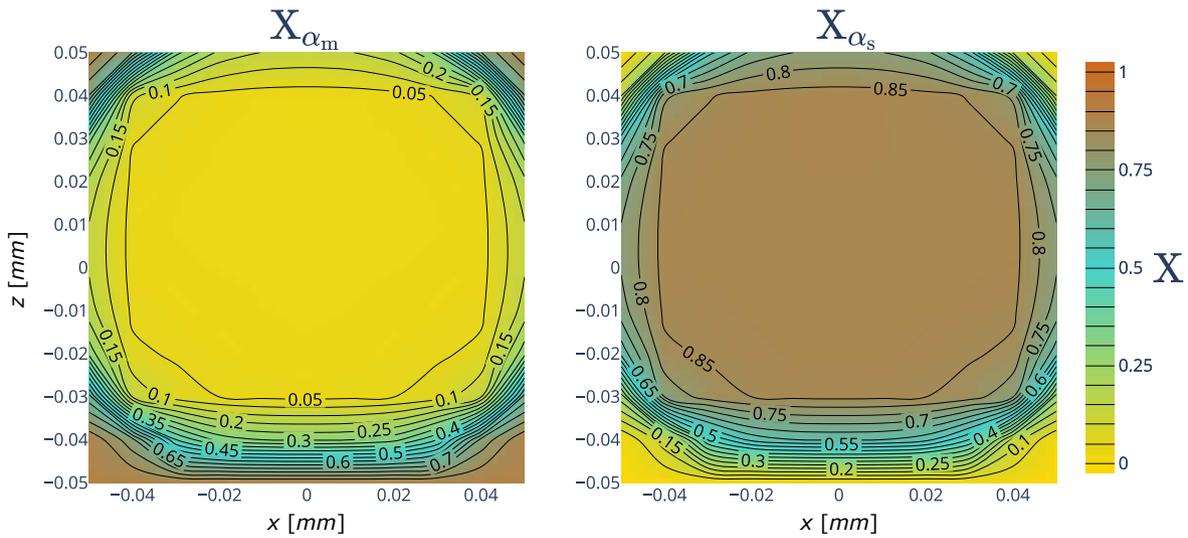}
    \caption{Simulated microstructure distribution of $\as$- and $\am$-phase in the vertical center plane of a ten centimeter-sided cube resulting from rapid cooling via thermo-mechanical contact at the bottom and free convection at the remaining surfaces~\cite{Meier2020_2}.}
    \label{fig: cube_Tbp_300}
 \end{figure}
 
In PBFAM parts of practically relevant size, the repeated thermal cycles in the process lead to temperature evolutions where material points in the center region of the part remain at elevated temperatures for considerable time spans such that also the creation of $\alpha_s$-phase, either directly out of the $\beta$-phase or via diffusion-based dissolution of Martensite at elevated temperatures (both effects are captured by the proposed model) can be expected. To investigate the influence of the increased thermal mass of larger parts on the centimeter scale, a final example was considered in~\cite{Meier2020_2}, where the quenching process of a cube with 100 mm side length was studied. The cube with a homogeneous initial temperature of $T_0=1300 \ K$ is in thermo-mechanical contact with a cold base-plate at its bottom surface (fixed temperature of $T_{\text{bp}}=300\ K$ at the bottom; effective heat-transfer coefficient of $\alpha_{\text{tc}}=5\cdot10^5 \frac{W}{m^2 K}$ at the (upper) contact interface with the cube) and subject to a convective boundary conditions (heat-transfer coefficient $\alpha_{\text{c}}=1000\ \frac{W}{m^2 K}$) on its remaining (free) surfaces. The microstructure evolution resulting from the rapid cooling induced by these boundary conditions is illustrated in Figure~\ref{fig: cube_Tbp_300}. 

Accordingly, a several millimeter thick Martensite coating at the surfaces is observed, which is especially pronounced at the corners of the cube that are characterized by higher cooling rates and lower temperature levels. These qualitative findings are in good agreement with experimental results for quenching of Ti-6Al-4V as well as additively manufactured parts~\cite{Kelly.Diss, kok2016geometry}. Due to the higher heat-transfer coefficient, a thicker Martensite coating is found at the bottom surface of the cube as compared to the remaining free surface areas. As expected, the core of the cube, where cooling rates are lower and the temperature remains at elevated levels for longer times, is composed of stable $\as$-phase. Moreover, in~\cite{Meier2020_2} it was demonstrated that heating of the base plate, either in the quenching or in the PBFAM example, is an effective means to suppress local Martensite formation. 

\vspace{-0.03 \textwidth}

{\subsection{Experimental validation}}
\label{micro_exp}

\noindent
{In addition to the experimental model validation via TTT- and CCT-diagrams as presented above, further steps of AM-related experimental microstructure characterization will be considered in our future research work. Well-established techniques for metallurgical characterization are typically used to identify the microstructural details of components made by LPBF and other metal AM processes. Optical microscopy of surfaces, typically etched to enhance the appearance of grain boundaries, is a routine method for identifying grain size and shape. Directional reflectance microscopy, combined with image processing, has been used for automated size segmentation~\cite{wittwer2021automated}.  Electron microscopy based methods such as electron backscatter diffraction (EBSD) are most commonly used to assess grain structure and orientation, again by surface analysis. Careful preparation of the sample surface for imaging, often by mechanical sectioning and polishing, is necessary and is commonplace in laboratories. Higher-resolution information, down to the nanoscale and atomic scale distribution of precipitates and alloying elements, can be obtained by transmission electron microscopy, atom probe tomography, and other techniques~\cite{wang2018additively}. These experimental methods are used to understand the relationships between process parameters, microstructure, and ultimate performance. For example, Kohnen and coworkers demonstrate how scan velocity influences EBSD-measured grain size and texture, and in turn the performance of tensile test artifacts fabricated via SLM~\cite{kohnen2019understanding} (also see, e.g.,~\cite{liu2019microstructure}).}\\

\section{Conclusion and outlook on overall modeling approach} \label{sec:wholistic}

\noindent
In this article, an overview was given on the authors' recent developments in the modeling of powder bed fusion additive manufacturing (PBFAM) processes on different length scales, specifically with respect to mesoscale powder modeling, mesoscale melt pool modeling, macroscale thermo-solid-mechanical modeling and microstructure modeling. Beyond the insights gained from these individual models, also their interplay in an integral modeling approach will be a central aspect of future research.

The long-term objective of the presented \textit{mesoscale} modeling approaches is to model the entire process chain from powder to part, i.e., the powder spreading as well as the subsequent laser melting and solidification process in PBFAM. First, the powder modeling approach according to Section~\ref{sec:powder} aims at the identification of key feedstock properties allowing to characterize the spreadability of new powder materials on basis of simple powder-rheological experiments. Second, the powder spreading simulations are intended to support the development of improved powder spreading strategies, e.g., novel spreading tools and kinematics, to enable controlled and repeatable spreading conditions that allow for high layer quality in terms of surface uniformity as well as high and spatially constant packing density. Specifically, a focus will lie on fine-grained, highly cohesive powders, for which a controlled spreading is challenging due to the dominating role of cohesive forces and thereby reduced flowability, but which are desirable for processes such as electron beam melting (EBM), selective laser melting (SLM) or binder jetting (BJ) for reasons of increased geometrical resolution, lower material costs and, in case of binder jetting, reduced sinter times. Taking the layers provided by the powder spreading model as input, the melt pool simulations of Section~\ref{sec:melt_pool_modeling} aim at correlating these layer properties with melt pool dynamics respectively instabilities, and finally with resulting defect characteristics, e.g., size, shape and frequency of pores. Specifically, the question shall be answered how melt pool stability can be fostered by improved opto-thermo-mechanical characteristics of high quality powder layers.

On the \textit{macroscale}, thermo-mechanical models according to Section~\ref{sec:macro_modeling} are intended to predict temperature fields, residual stresses and thermal distortion on the scale of design parts. The predicted temperature fields can be taken as input to predict metallurgical phase fraction evolutions using (homogenized) \textit{microstructure} models as presented in Section~\ref{sec:micro_modeling}. The knowledge of these phase fractions can be exploited to formulate microstructure-informed, macroscale constitutive laws, which are calculated by spatially homogenizing the material properties of the individual species, and which allow for more accurate residual stress predictions when integrated into thermo-mechanical PBFAM models. Due to the extreme spatial and temporal temperature gradients during PBFAM, which fosters the formation of non-equilibrium (e.g., martensitic) phases, and the considerably different constitutive behavior of these diverse phases (e.g., ductility of $\beta$-phase vs. martensitic $\alpha_m$-phase in Ti-6Al-4V), this approach is expected to significantly increase model accuracy.

Also an information transfer between \textit{macro}- and \textit{mesoscale} can be beneficial for overall modeling. On the one hand, macroscale models can provide improved thermal boundary conditions for representative volumes at different locations of the design part, which are considered for mesoscale melt pool simulations. On the other hand, the mesoscale powder and melt pool models may be exploited to derive improved effective continuum properties (e.g., effective opto-thermal properties of powder phase, anisotropic thermal conductivity as model for convective heat transfer in the melt, etc.) for the macroscale model.

{Many of the parameters underlying the aforementioned models cannot be taken from standard databases due to the extreme processing conditions (e.g., extreme spatial and temporal temperature gradients) in PBFAM and the high sensitivity/variability of certain model parameters with respect to physical state variables (e.g., temperature, pressure, etc.), environmental conditions and material imperfections. In our ongoing research, high-resolution measurement data from powder spreading and laser melting experiments as described in Sections~\ref{powder_exp} and~\ref{melt_exp} will be combined with probabilistic schemes for inverse parameter identification and uncertainty quantification to identify the unknown model parameters under process-relevant conditions. {In addition to parameter identification/calibration, the underlying methods for inverse analysis will also be used to directly (i.e., without repeated and computationally expensive forward simulations in a 'trial-and-error' manner) answer practically relevant inverse questions, i.e., the question which input is required to achieve a desirable output. An example in this context is the geometrical compensation of thermal distortion by solving the inverse design task of finding an optimal initial geometry}}

Given the multiscale nature of metal PBFAM processes and the complexity arising from competing physical mechanismen and various types of defects on these different length scales, an integrated modeling and simulation approach unifying information from macro-, meso- and microscale models has the potential for virtual process optimization and part qualification, thus promoting broader industrial adoption.

\section*{Acknowledgments}
The authors would like to thank Gareth McKinley and Crystal Owens from MIT, Marvin Ochsenius from TUM as well as Jonah Myerberg from Desktop Metal for fruitful discussions and their support in the context of powder-rheological experiments. {Further, the authors would like to thank their TUM colleagues Peter Munch for his valuable support in the code implementation of the (FEM) melt pool model, Niklas Fehn for his valuable contributions to the DEM code implementation, and Martin Kronbichler for fruitful discussions in the context of DEM as well as SPH and FEM (melt pool model) code implementations.} C. Meier acknowledges funding by a postdoc fellowship of the German Academic Exchange Service (DAAD). Further, the TUM authors acknowledge funding of the Deutsche Forschungsgemeinschaft (DFG, German Research Foundation) within project 437616465 and project 414180263. {Magdalena Schreter was supported by the Austrian Science Fund (FWF) via a FWF Schr\"odinger scholarship, FWF project number J-4577-N. Y. Sun acknowledges funding by the CSC scholarship with number 201909110.} For the cited experimental work on X-ray microscopy of powder layers and MWIR microscopy of LPBF, the MIT authors acknowledge prior and current financial support by Robert Bosch, LLC; Honeywell Federal Manufacturing and Technologies (FM \& T); and ArcelorMittal.

{\section*{Authors' contributions}}
{
All authors contributed to writing and proof-reading the manuscript. C. Meier contributed to the conceptualization and derivation of model equations w.r.t. the different modeling approaches, as well as to the code implementation and conduction of numerical simulations w.r.t. the powder and (SPH) melt pool model. S.L. Fuchs contributed to the conceptualization, derivation of model equations, the code implementation and conduction of numerical simulations w.r.t. the powder and (SPH) melt pool model. N. Much contributed to the code implementation of the macroscale and (FEM) melt pool model. J. Nitzler contributed to the derivation of model equations, the code implementation and conduction of numerical simulations w.r.t. the microstructure model. R.W. Penny contributed to the conceptualization, design and conduction of powder spreading experiments including x-ray metrology. P.M. Praegla contributed to the derivation of model equations, the code implementation and conduction of numerical simulations w.r.t. the powder model. S.D. Pr\"oll contributed to the derivation of model equations, the code implementation and conduction of numerical simulations w.r.t. the macroscale model. Y. Sun contributed to the code implementation and conduction of numerical simulations w.r.t. the (SPH) melt pool model. R. Weissbach contributed to the derivation of model equations, the code implementation and conduction of numerical simulations w.r.t. the powder model. M. Schreter contributed to the conceptualization, derivation of model equations, the code implementation and conduction of numerical simulations w.r.t. the (FEM) melt pool model. N. Hodge contributed to the conceptualization, derivation of model equations, the code implementation and conduction of numerical simulations w.r.t. the microstructure model. A.J. Hart contributed to the conceptualization and design of the different experimental approaches, and supervised work at MIT. W.A. Wall contributed to the conceptualization of the different modeling approaches, and supervised work at TUM.}\\

\bibliography{collection.bib}

\end{document}